\documentclass[traditabstract]{aa}  

\usepackage{graphicx}

\usepackage{txfonts}
\usepackage{natbib}
\begin{document}

   \title{Broadband X-ray observations of four gamma-ray narrow-line Seyfert 1 galaxies}


   \author{M. Berton\inst{1,2}\thanks{e-mail: marco.berton@utu.fi}
          \and
          V. Braito\inst{3}\and
	  S. Mathur\inst{4}\and
	  L. Foschini\inst{3}\and
	  E. Piconcelli\inst{5}\and
	  S. Chen\inst{6,7,8}\and
	  R. W. Pogge\inst{4}
          }

   \institute{
	$^1$ Finnish Centre for Astronomy with ESO (FINCA), University of Turku, Quantum, Vesilinnantie 5, FI-20014 University of Turku, Finland; \\
	$^2$ Aalto University Mets{\"a}hovi Radio Observatory, Mets{\"a}hovintie 114, FI-02540 Kylm{\"a}l{\"a}, Finland; \\
	$^3$ INAF - Osservatorio Astronomico di Brera, via E. Bianchi 46, 23807 Merate (LC), Italy; \\
	$^4$ Department of Astronomy and Center for Cosmology and AstroParticle Physics, The Ohio State University, 140 West 18th Avenue, Columbus, OH 43210, USA; \\
	$^5$ Osservatorio Astronomico di Roma (INAF), via Frascati 33, I-00040 Monte Porzio Catone (Roma), Italy \\
 	$^6$ Dipartimento di Fisica e Astronomia "G. Galilei", Universit\`a di Padova, Vicolo dell'Osservatorio 3, 35122 Padova, Italy \\
	$^7$ Center for Astrophysics, Guangzhou University, 510006, Guangzhou, China \\
	$^8$ Istituto Nazionale di Fisica Nucleare (INFN), Sezione di Padova, 35131, Padova, Italy \\
             }

   \date{}

\authorrunning{M. Berton et al.}
\titlerunning{Broad-band X-ray observations of four $\gamma$-NLS1s}
 
  \abstract{Narrow-line Seyfert 1 galaxies (NLS1s) is one of the few classes of active galactic nuclei (AGN) harboring powerful relativistic jets and detected in $\gamma$ rays. NLS1s are well-known X-ray sources. While in non-jetted sources the origin of this X-ray emission may be a hot corona surrounding the accretion disk, in jetted objects, especially beamed ones, the contribution of corona and relativistic jet is difficult to disentangle without a proper sampling of the hard X-ray emission. For this reason, we observed with \textit{NuSTAR} the first four NLS1s detected at high energy $\gamma$ rays. These data, along with \textit{XMM-Newton} and \textit{Swift/XRT} observations, confirmed that X rays originate both in the jet and in the accretion disk corona. Time variability in hard X rays furthermore suggests that, as observed in flat-spectrum radio quasars, the dissipation region during flares could change its position from source to source, and it can be located both inside and outside the broad-line region. We find that jetted NLS1s, and other blazars as well, seem not to follow the classical fundamental plane of BH activity, which therefore should be used as a BH mass estimator in blazars with extreme care only. Our results strengthen the idea according to which $\gamma$-NLS1s are smaller and younger version of flat-spectrum radio quasars, in which both a Seyfert and a blazar component co-exist. }


   \authorrunning{M. Berton et al.}

   \maketitle
%

\section{Introduction}

Narrow-Line Seyfert 1s (NLS1s) are a subclass of Seyfert 1 galaxies (Sy1s) defined by relatively narrow permitted emission lines (full width at half maximum (FWHM) $<$ 2000 km s$^{-1}$). 
They often show strong optical Fe II emission relative to H$\beta$, and weak [O III] emission \citep{Osterbrock85, Goodrich89}. 
These objects started out as little more than curiosities. 
However, since their first classification, the extreme properties of NLS1s have made them important objects to probe accretion processes in active galactic nuclei (AGN). 
Furthermore, they occupy one extreme end of the Boroson \& Green Principal Component 1 (PC1, also called Eigenvector 1, EV1; \citealp{Boroson92}). 
The physical driver behind EV1 is debated, but is usually thought to correlate with the luminosity relative to the Eddington luminosity, known as the Eddington ratio, although inclination may play a role \citep{Shen14}. 
The NLS1s lie at the extreme end of EV1 corresponding to high Eddington ratio \citep{Pounds95, Sulentic00, Marziani18a}, and also seem to have lower BH (BH) masses than broad-line AGNs of similar luminosity (e.g., \citealp{Peterson11}). 
As such, NLS1s allow us to probe the higher end of the accretion mechanisms. \par
Like normal Sy1s, NLS1s are mostly radio-quiet. 
The discovery of some radio-loud NLS1s (RLNLS1) therefore came as a surprise \citep{Grupe00, Zhou03}, given their smaller supermassive BHs (SMBHs) and the expectation that radio-loudness is correlated with BH mass \citep[e.g.,][]{Chiaberge11}. 
Even more surprising has been the detection of a subset of NLS1s in $\gamma$ rays with the \textit{Fermi} Gamma-ray Space Telescope \citep{Abdo09a,Abdo09c}. 
This conclusively confirmed the presence of powerful beamed relativistic jet in RLNLS1s. 
To date, 20 NLS1s are known to be sources of $\gamma$-rays at least during their flaring activity \citep[for a full list, see][and references therein]{Romano18}. 
The detection at very high energies of these flat-spectrum RLNLS1s (F-NLS1) shook the paradigm according to which only AGN hosted in elliptical galaxies with large BH mass were able to launch powerful relativistic jets \citep{Laor00}. 
NLS1s are indeed hosted in disk galaxies \citep{Deo06, OrbandeXivry11, Mathur12}, and F-NLS1s may not be different \citep{Anton08, Kotilainen16, OlguinIglesias17, Jarvela18, Berton19a}. \par
A widely known interpretation of the nature of NLS1s is that they represent a young phase of AGN activity, eventually evolving into Sy1s \citep{Mathur00, Mathur01}. 
The same is thought to be the case for F-NLS1s. 
These latter be a young evolutionary phase of blazars, particularly of flat-spectrum radio quasars (FSRQs), since they constitute the low-mass and low-luminosity tail of FSRQs distribution \citep{Abdo09a, Abdo09c, Foschini11, Foschini15, Berton16c, Berton17, Berton18a}. \par
One of the characterizing properties of NLS1s is that they are strong soft X-ray emitters, and constitute almost half of the AGN detected between 0.1 and 2 keV by the \textit{ROSAT} All Sky Survey \citep{Grupe96}. 
Their X-ray spectrum usually exhibits a soft excess below 2 keV with a steep power law at higher energies up to 10 keV ($\Gamma \sim 2.5$, \citealp{Brandt97}), possibly associated to the presence of the thermally Comptonized X-ray corona and the accretion disk \citep[e.g.,][]{Grupe04, Done12, Gallo18}.
The spectrum also shows enhanced variability with respect to broad-line Seyfert 1s (BLS1s), further possible evidence that the BH mass of these objects is typically smaller than in their broad-line counterparts \citep{Leighly99a}. 
In the 20-100 keV band the photon index can be extremely steep, down to values of around three, but on average comparable to BLS1s in the same spectral region \citep{Dadina07, Dadina08, Panessa11}. 
These hard-X-ray-selected NLS1s also show a less prominent soft excess with respect to soft-X-ray-selected NLS1s and a lower Eddington ratio, another property reminiscent of BLS1s \citep{Bianchi09, Grupe10}. \par
F-NLS1s have a bimodal distribution of photon indexes, due to different states of jet activity.
In \citet{Foschini15}, 23 out of 42 F-NLS1s were shown to have hard photon index ($\Gamma <$ 2), like hard-X-ray-selected NLS1s and similarly to regular Sy1s ($\Gamma \sim$ 1.7-2.0, \citealp{Reeves00, Caccianiga04, Piconcelli05}). 
Others, instead, are reminiscent of typical soft-X-ray-selected radio-quiet NLS1s (RQNLS1, $\Gamma \sim$ 2.7).
Their Eddington ratio however is not different from that of regular NLS1s \citep{Berton15a, Rakshit17a, Chen18}.
This property, along with their spectral energy distributions (SEDs) and the bimodal distribution of their photon indexes, indicates that the origin of their X-ray does not reside only in the corona, but that the jet contribution is extremely important \citep{Foschini15}, as observed in other jetted AGN \citep{Leighly99b, Piconcelli05}. 
Like in other beamed and jetted sources, particularly FSRQs, the X- and $\gamma$-ray region of their SED is dominated by inverse Compton (IC), in particular external Compton (EC) \citep[e.g.,][]{Landt08}. 
Therefore, F-NLS1s may be objects where both a Seyfert (accretion disk or corona) and a blazar (relativistic jet) component are co-existing. \par
Observations of jetted NLS1s however have been carried out mostly below 10 keV. 
The X-ray properties of beamed jetted NLS1s above this energy threshold have never been studied in detail, although this part of the spectrum is essential to disentagle the contribution from the jet and the corona, otherwise a challenging accomplishment, and to establish whether the hard X rays of $\gamma$-NLS1s originate in the corona, in the jet, or in both.
The Nuclear Spectroscopic Telescope Array (\textit{NuSTAR}, \citealp{Harrison13}) can provide new insights to this relatively unexplored spectral region. 
In this work we study the properties of the first four NLS1s detected in $\gamma$ rays observed with \textit{NuSTAR}, 1H 0323+342, PMN J0948+0022, PKS 1502+036, and PKS 2004-447 \citep{Abdo09c}. 
We present here a detailed analysis of their X-ray spectra combining \textit{NuSTAR} observations with those of other X-ray satellites, and light curves in the hard X-rays. \par
In Sect. 2 we provide a brief general introduction on the properties of the four $\gamma$ ray NLS1s, in Sect. 3 we describe the \textit{NuSTAR}, \textit{XMM-Newton}, and \textit{Swift/XRT} observations and data analysis, in Sect. 4 we present the spectral analysis source by source and compare our results with the literature, in Sect. 5 we study the time variability in the \textit{NuSTAR} light curve, in Sect. 6 we discuss the position of our sources in the fundamental plane of BH activity \citep{Merloni03}, in Sect. 7 we discuss how NLS1s can be included in the blazar sequence, and in Sect. 8 we summarize our conclusions.  
In the following, we adopt the standard $\Lambda$CDM cosmology, with H$ = 70$ km s$^{-1}$ Mpc$^{-1}$, $\Omega_m = 0.3$, $\Omega_\Lambda = 0.7$ \citep{Komatsu11}.

\section{Target description}
\begin{table}
\caption{List of $\gamma$-NLS1s observed by \textit{NuSTAR}.}
\label{tab:source_list}
\centering
\footnotesize
\scalebox{0.85}{
\begin{tabular}{l c c c c c}
\hline\hline
Name & Alias & R.A. & Dec. & z  & F$_r$\\
\hline
1H 0323+342 & J0324 & 51.17151 & $+$34.17940 & 0.061 & 304$^a$ \\
PMN J0948+0022 & J0948 & 147.23883 & $+$0.37377 & 0.585 & 305$^b$ \\
PKS 1502+036 & J1505 & 226.27699 & $+$3.44189 & 0.407 & 403$^c$ \\
PKS 2004-447 & J2007 & 301.97993 & $-$44.57897 & 0.240 & 446$^d$ \\
\hline\hline
\end{tabular}
}
\tablefoot{Columns: (1) name; (2) short name; (3) right ascension (degrees, J2000); (4) declination (degrees, J2000); (5) redshift; (6) radio flux density (mJy) at 5 GHz (references $^a$\citealp{Laurent97}, $^b$\citealp{Doi06}, $^c$\citealp{Berton18a}, $^d$\citealp{Gallo06a}).}
\end{table}
The four sources studied in this paper are the four NLS1s with the highest flux in $\gamma$ rays, and were all discovered as high-energy sources after the first year of operations of the \textit{Fermi} satellite in 2008 \citep{Abdo09c}. The source list is presented in Table \ref{tab:source_list}. \par
\textbf{1H 0323+342 = J0324} is the closest $\gamma$ NLS1. Hosted by a spiral galaxy or an interacting late-type galaxy \citep{Zhou07, Anton08, Leontavares14} with a BH mass from reverberation mapping of 3.4$\times10^7$ M$_\odot$ \citep{Wang16}, this NLS1 has the lowest $\gamma$ ray luminosity of the sample, but because of its short distance it was detected by \textit{Fermi} soon after its launch \citep{Abdo09c}. This object was detected in hard X-rays for the first time by \textit{INTEGRAL} \citep{Bird07, Malizia07}, but it was classified as a regular Seyfert. Only later was its spectral variability revealed, with a low flux and steep spectrum in the \textit{INTEGRAL} observation, and high flux and hard spectrum in a following \textit{Swift/Burst Alert telescope} \citep{Barthelmy05} observation, suggesting the presence of a relativistic jet \citep{Foschini09}. J0324 is the only $\gamma$-NLS1 showing a Fe K$\alpha$ emission line \citep{Abdo09c, Kynoch18}. 
 \par
\textbf{PMN J0948+0022 = J0948} is the first NLS1 ever detected in $\gamma$ rays, and the most luminous one at very high energies \citep{Abdo09a}. Identified as a radio-loud NLS1 immediately after the release of the Sloan Digital Sky Survey (SDSS, \citealp{Zhou02}), its BH mass of 7.5$\times$10$^7$ M$_\odot$ lies close to the divide between NLS1s and FSRQs \citep{Foschini15}. We remark that all BH masses from \citet{Foschini15} (J0948, J1505, J2007) were calculated using the virial theorem. As a proxy for velocity these latter authors used the second-order moment of the broad component of H$\beta$ line \citep{Peterson04}, while the broad-line region (BLR) radius was calculated assuming it was proportional to the H$\beta$ luminosity \citep{Greene10}. The $f$ factor was derived by \citet{Collin06}. The host galaxy, as in many other NLS1s, is a disk galaxy with a pseudobulge \citep{Jarvela18}. This source is well known for its strong variability at all frequencies \citep{Abdo09b, Liu10, Foschini12}, with repeated flaring activity \citep{Foschini11c, Dammando15a}. The X-ray spectrum is also variable, with a photon index $\Gamma$ ranging from 1.3 to 1.8 \citep{Foschini15}, and it is characterized by a soft excess below 2.5 keV, which can be associated with thermal Comptonization, and possibly by a power law at higher energies, produced by the relativistic jet \citep{Bhattacharyya14}. \par
\textbf{PKS 1502+036 = J1505} has a BH mass comparable to that of J0324 (1.9$\times$10$^7$ M$_\odot$, \citealp{Foschini15}), but its $\gamma$-ray luminosity is the second highest among our NLS1s \citep{Abdo09c}. In radio it shows a core-jet structure both at parsec \citep{Orienti12} and kiloparsec scale \citep{Berton18a}, and it may be the only source in our sample hosted by an elliptical galaxy \citep[][]{Dammando18}. In X-rays its spectrum is typically well reproduced by a single power law, although a possible break may be present at higher energies \citep{Foschini15}. The X-ray spectrum became extremely hard ($\Gamma \sim 1$, \citealp{Dammando16}) during its flaring activity \citep{Paliya16}, a harder-when-brighter behavior similar to that often observed in blazars \citep[e.g.,][]{Giommi90, Kalita17, Berton18b}. \par
\textbf{PKS 2004-447 = J2007} is the fist $\gamma$ NLS1 identified in the southern hemisphere. Its BH mass and Eddington ratio are typical for a NLS1 (7.0$\times$10$^7$ M$_\odot$ and 0.05, respectively \citealp{Foschini15}, but its other properties are peculiar with respect to the bulk of the NLS1 population. Its radio properties closely indeed remind those of compact steep-spectrum sources \citep{Oshlack01, Schulz15}. Like J0324, it is hosted in a late-type galaxy with a pseudobulge \citep{Kotilainen16}. Its blazar-like behavior was noted already by \citet{Gallo06a}, and although a single power law usually dominates the X-ray spectrum with a typical photon index of 1.6 \citep{Kreikenbohm16}, in some cases it can also show a soft excess attributed to the corona \citep{Foschini09}. \par

\section{Observations and data reduction}
\subsection{NuSTAR}
All the NuSTAR observations were reduced using the standard \texttt{nupipeline v0.4.6}, the \texttt{HeaSoft v6.25}, along with the \texttt{caldb v2018-10-30}. 
The details of NuSTAR observations are reported in Table~\ref{tab:obs_nustar}. 
Light curves and spectra were extracted separately for both focal plane modules FPMA and FPMB, using in all cases a circular aperture centered on the source coordinates and with a radius of 60$^{\prime\prime}$. 
The background was extracted out of a radius of 90 arcsecs on the same chip where the source was located. 
The spectra were binned differently according to the total number of counts, with 50 counts per bin for J0948, J1505, and J2007, and 100 counts per bin for J0324.
For all sources we also reproduced light curves between 4 and 50 keV, co-adding the FPMA and FPMB light curves, and using a time binning of 5814 s, corresponding to a NuSTAR orbit. \par

\subsection{XMM-Newton}
Both J0948 and J2007 were observed with \textit{XMM-Newton} both with the European Photon Imaging Camera (EPIC) pn \citep{Struder01} and MOS detectors \citep{Turner01}.
For J0948 the observation was performed simultaneously with \textit{NuSTAR}.
For J2007, an \textit{XMM} observation was carried out four days apart from one of the \textit{NuSTAR} observations. 
The other \textit{NuSTAR} observation does not include any quasi-simultaneous low-energy observations. 
We processed and cleaned each of the observations using the \textit{XMM-Newton} Science Analysis Software (SAS ver. 16.0.0 ) and the latest available calibration files. 
The observation of J0948 was only marginally affected by background flaring, while in J2007 a period of background flaring occurred towards the end of the observation. 
In the case of J0948 the resulting net exposure times are 63.9, 84.9, and 89.5 ks for the pn, MOS1, and MOS2, respectively. 
For J2007 the net exposure times are 31.5, 39.7, and 39.6 ks for the pn, MOS1, and MOS2, respectively. 
Response matrices and ancillary response files at the source position were created using the SAS tasks {\it arfgen} and {\it rmfgen}. 
For the EPIC-pn spectra we used single- and double-pixel events (i.e., pattern $<=$ 4), while for the EPIC-MOS spectra we used the standard pattern $<=12$ which includes also the triple and quadruple events. 
For J0948, source and background spectra were extracted for each of the detectors using a circular region with a radius of 29$^{\prime\prime}$ and two circular regions with the same radius. 
The source spectra were then binned to have at least 50 counts in each energy bin. 
For J2007, the source spectra were extracted adopting a circular region  with a radius of 29$^{\prime\prime}$, while the background spectra were extracted from two circular regions, each with radius of 30$^{\prime\prime}$. \par
In the case of J2007 spectra, to avoid problems due to nonsimultaneous observation, we tested whether the photon index remained approximately constant in the common energy range between NuSTAR and XMM. 
Since all of them are consistent, we proceeded further with the analysis of all the spectra together.

\subsection{Swift/XRT}
For J1505, one simultaneous observation to \textit{NuSTAR} was performed with the X-Ray Telescope (\textit{XRT}, \citealp{Burrows05}) on board the \textit{Swift} satellite.
The data were reduced using the standard approach with \texttt{xrtpipeline v0.13.4}, but the number of counts was too low to apply the same $\chi^2$ analysis carried out for \textit{NuSTAR}. 
Furthermore, for J0324 no simultaneous soft X-ray observations were available at all. 
For this reason, to have a good sampling in soft X-rays, we decided to add together the ten available \textit{XRT} observations closest in time, to improve our statistics by one order of magnitude.
As before, we tested whether the photon index was constant in the common energy range. 
While for J1505 the spectra turned out to be consistent, for J0324 we were not able to find any observation in which the photon index was the same. 
Therefore, we proceeded without the 0.3-3 keV spectrum for J0324. 
The fluxes in this interval range were estimated by extrapolating the spectrum.
For J1505, the total exposure time with XRT is 20 ks.
The observational details are shown in Table~\ref{tab:obs_swift}.  \par

\begin{table}[!t]
\caption{Spectral fitting parameters for J0324.}
\centering
\footnotesize
\begin{tabular}{l l}
\hline
\multicolumn{2}{l}{\textbf{TBabs*bknpo}} \\
\hline\hline
nH & 1.27$\times10^{21}$ \\
$\Gamma_1$ & 1.83$\pm$0.02 \\
E$_{\rm break}$ & 13.40$^{+3.39}_{-3.86}$ \\
$\Gamma_2$ & 1.68$\pm$0.07 \\
norm$_1$ & (2.50$\pm$0.11)$\times 10^{-3}$ \\
c$_{FPMB}$ & 1.05$\pm$0.02 \\
$\chi^2_\nu$/dof & 1.11/316 \\
Flux$^o_{0.3-70}$ & 27.8$\pm$1.5 \\
logL$^i_{0.3-70}$ & 44.38$\pm$0.06 \\
\hline\hline
\end{tabular}
\tablefoot{Lines: (1) Galactic hydrogen column density from \citet{Kalberla05} (cm$^{-2}$); (2) photon index below the break energy; (3) break energy (keV); (4) photon index above the break energy; (5) normalization of the spectrum; (6) cross-calibration constants for FPMB; (7) $\chi^2_\nu$ and degrees of freedom; (8) observed flux between 0.3 and 70 keV in units of $10^{-12}$ erg s$^{-1}$ cm$^{-2}$; (9) logarithm of the intrinsic luminosity between 0.3 and 70 keV.}
\label{tab:J0324}
\end{table}

\section{Spectral analysis}
The spectral analysis was performed with \texttt{XSPEC v12.10.1}. 
The spectra and the best model we used to reproduce them are shown in Fig.\ref{fig:spectra}. 
We fitted all the available data simultaneously for each source, that is the low-energy part from \textit{XMM-Newton} pn and MOS or \textit{XRT}, and the high-energy part from \textit{NuSTAR} FPMA and FPMB.  
We note that due to nonsimultaneous observations, there is some variability between different spectral regions observed with different satellites. 
This could affect our measurements, especially those of the photon indexes, which therefore should be taken with care. 
For each source, we modeled the spectra starting with a redshifted single power law. 
Given that in the past some sources showed an energy break in their soft X-ray spectra \citep{Foschini15}, we also tested in all cases a broken power-law model. 
The latter is the phenomenological representation of a two-component spectrum. 
The spectral region below the energy break may be associated with coronal emission, while above the break the relativistic jet may be dominant.
To provide a more physical model we used CompTT, an analytic model which describes the Comptonization of soft photons in a hot plasma \citep{Titarchuk94}, on top of a single redshifted power law representing the relativistic jet continuum.  
Finally, since the spectrum may show a cut-off due to pair production, we modeled the spectra with a cut-off power law. 
All models were combined with a Tuebingen-Bolder of interstellar medium absorption model to represent Galactic absorption. 
We used the hydrogen column density (nH) values derived by \cite{Kalberla05}. \par
In addition, for the three sources with the most counts we decided to test some additional models. 
Specifically, we also tried to combine the underlying power law with a black body emission, a warm absorber model, and finally a Gaussian component to reproduce the Fe K$\alpha$ line at 6.4 keV.
The results of the spectral fitting are shown in Table \ref{tab:results}, and are described in the following section. 
The observed fluxes and the intrinsic luminosities calculated with different spectral models are shown in Table \ref{tab:fluxes}. \par

\subsection{J0324}

\begin{figure*}[htb]
    \centering
    \begin{minipage}[t]{0.8\textwidth}
        \centering
        \includegraphics[width=\textwidth]{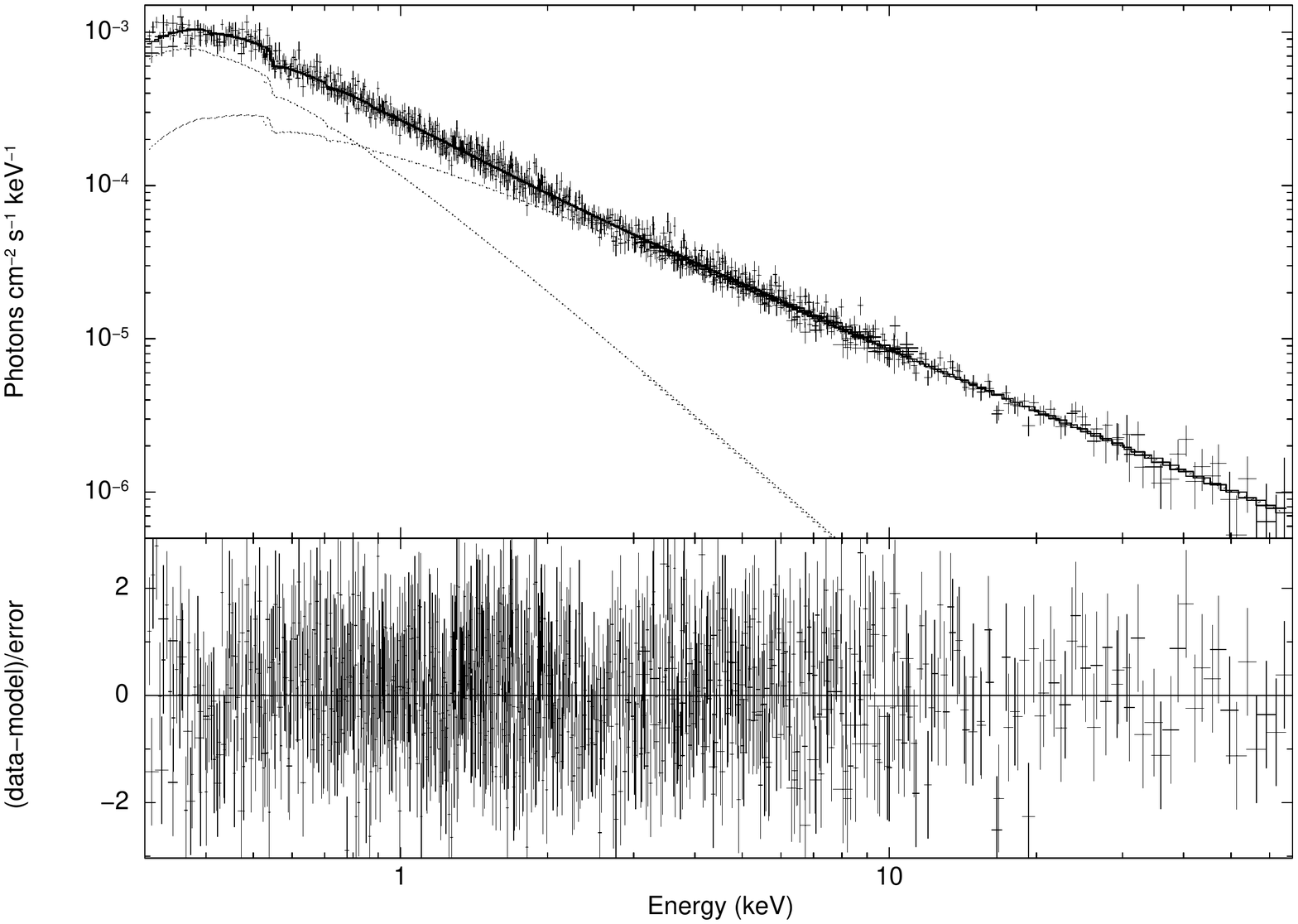}
    \end{minipage}
    \hfill
    \begin{minipage}[t]{.33\textwidth}
	\centering
	\includegraphics[width=\textwidth]{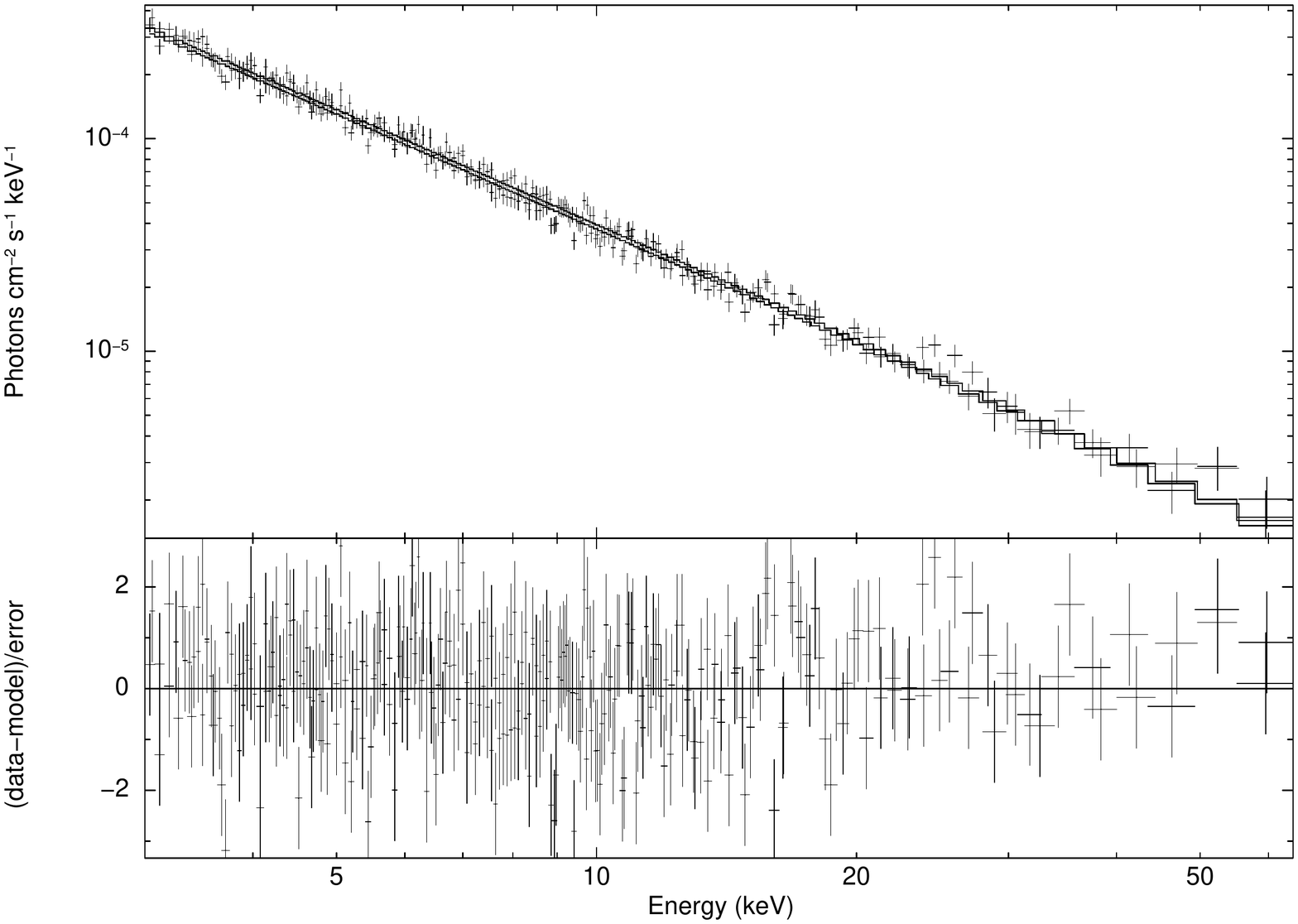}
    \end{minipage}
    \hfill
    \begin{minipage}[t]{.33\textwidth}
        \centering
        \includegraphics[width=\textwidth]{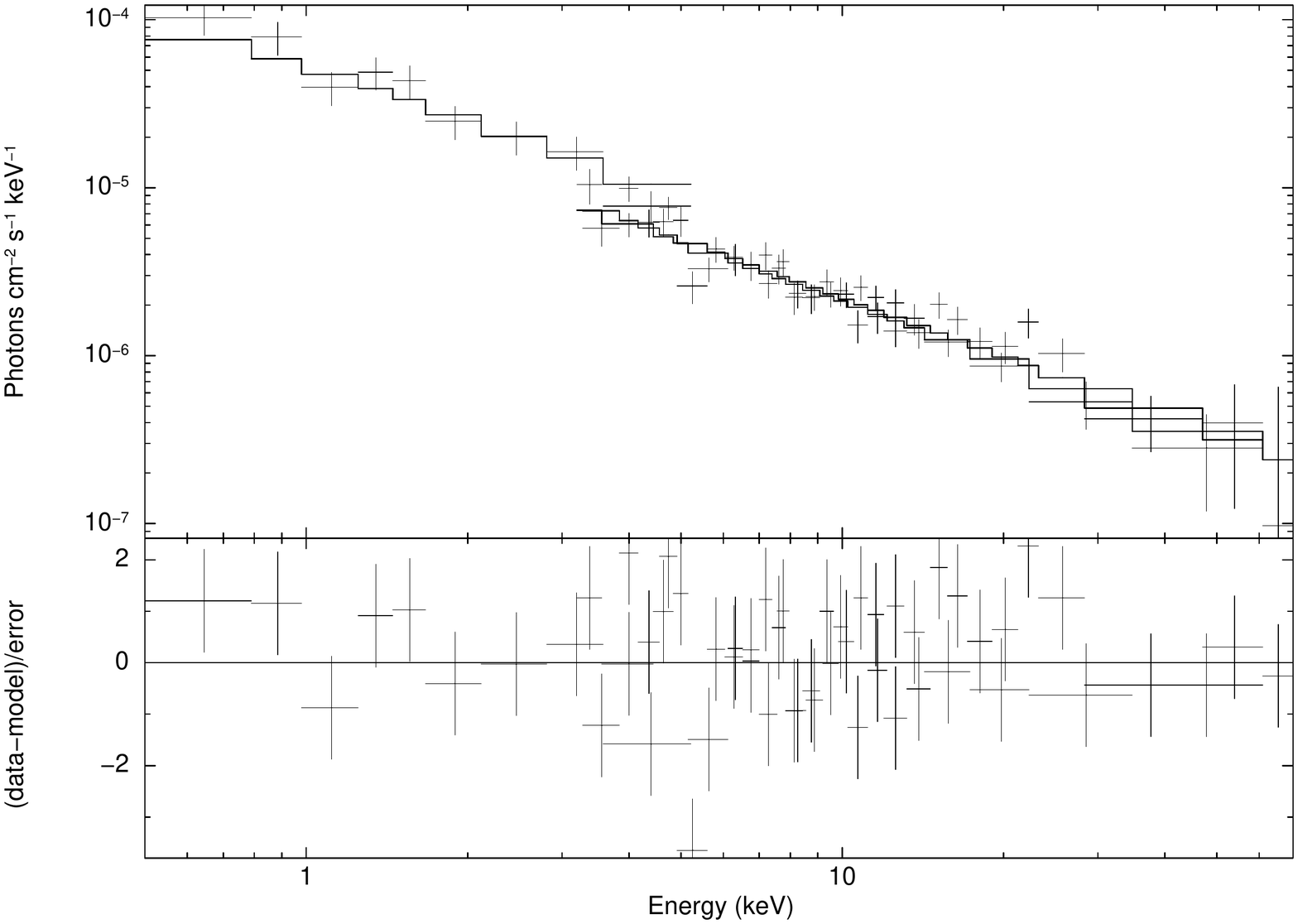}
    \end{minipage} 
    \hfill
    \begin{minipage}[t]{.33\textwidth}
        \centering
        \includegraphics[width=\textwidth]{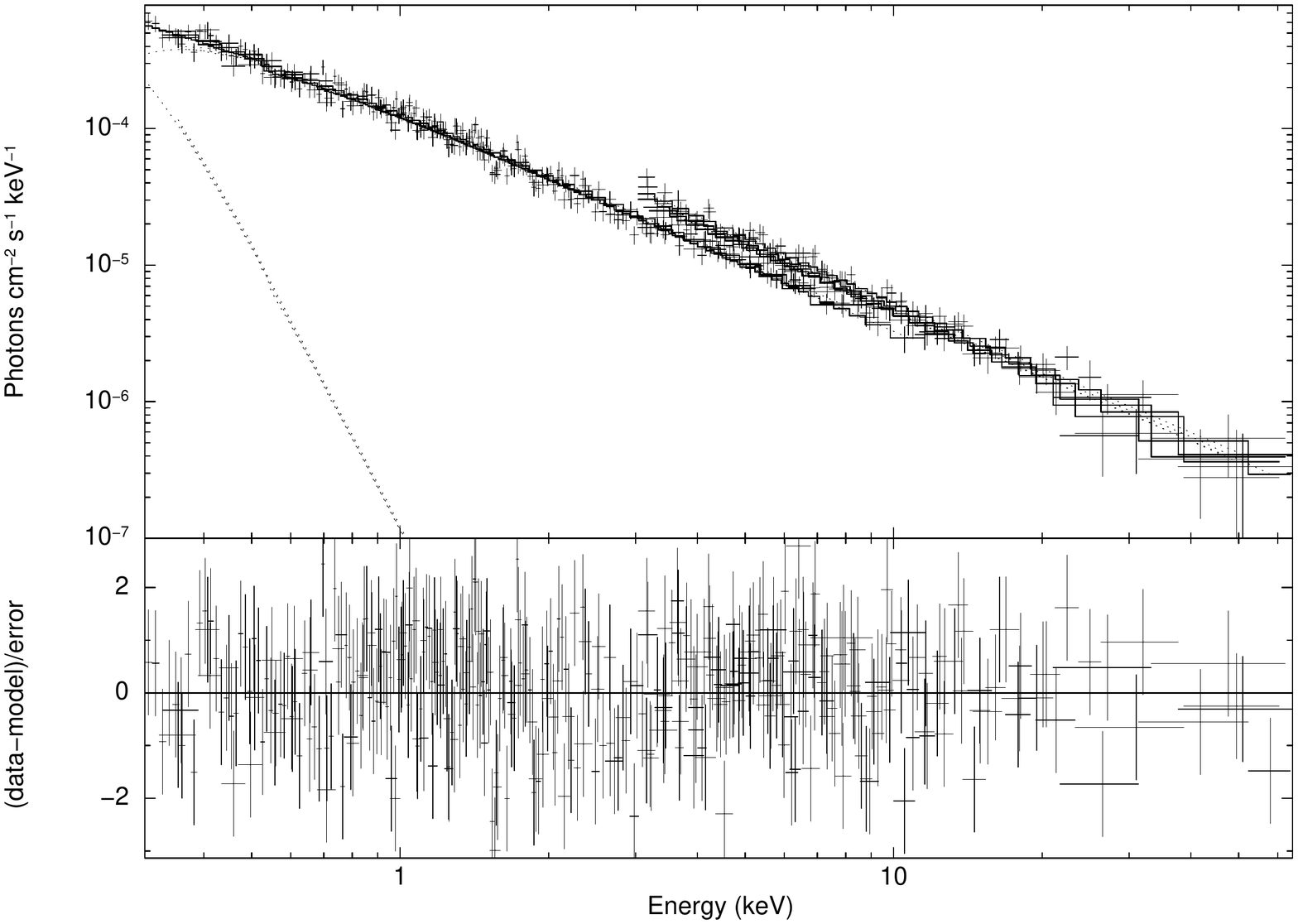}
    \end{minipage}  
\caption{Spectra of J0948 (top row), and in the bottom row from left to right J0324, J1505, and J2007. The spectra of J0948 and J2007 were modeled with Galactic absorption, thermal comptonization, and a redshifted power law. The spectrum of J1505 was modeled with Galactic absorption and a single power law, and that of J0324 with Galactic absorption and a broken power law. The offsets visible in J1505 and J2007 are due to nonsimultaneous observations of different satellites. }
\label{fig:spectra}
\end{figure*}

\begin{table}
\caption{Spectral fitting parameters for J0948.}
\centering
\footnotesize
\begin{tabular}{l l}
\hline
\multicolumn{2}{l}{\textbf{TBabs*(zpo+compTT)}} \\
\hline\hline
nH & 0.55$\times10^{21}$ \\
$\Gamma$ & 1.31$\pm$0.04 \\
norm$_1$ & (3.19$^{+0.33}_{-0.17}$)$\times 10^{-4}$ \\
T$_0$ & $\leq$ 81 \\
kT & $\geq$2.1 \\
$\tau_p$ & $\leq$5.2 \\
norm$_2$ & $\leq$3.6$\times 10^{-2}$ \\
c$_{MOS1}$ & 1.02$\pm$0.02 \\
c$_{MOS2}$ & 1.01$\pm$0.02 \\
c$_{FPMA}$ & 0.96$\pm$0.02 \\
c$_{FPMB}$ & 0.98$\pm$0.02 \\
$\chi^2_\nu$/dof & 1.01/1218 \\
Flux$^o_{0.3-70}$ & 7.74$\pm$0.77 \\
logL$^i_{0.3-70}$ & 45.79$\pm$0.10 \\
\hline\hline
\end{tabular}
\tablefoot{Lines: (1) Galactic hydrogen column density from \citet{Kalberla05} (cm$^{-2}$); (2) photon index of the redshifted power law; (3) normalization of the power law component; (4) seed photon temperature (eV); (5) plasma temperature (keV); (6) optical depth of the plasma; (7) normalization of thermal Comptonization model; (8),(9),(10),(11) cross-calibration constants for FPMA and FPMB; (10) $\chi^2_\nu$ and degrees of freedom; (12) observed flux between 0.3 and 70 keV in units of $10^{-12}$ erg s$^{-1}$ cm$^{-2}$; (13) logarithm of the intrinsic luminosity between 0.3 and 70 keV.}
\label{tab:J0948}
\end{table}
As mentioned above, J0324 is the only $\gamma$-NLS1 in which the Fe K$\alpha$ emission line has been detected in both the \textit{Swift/XRT}, \textit{XMM}, and \textit{Suzaku} data \citep{Abdo09c, Kynoch18, Paliya19}. 
As in \citet{Landt17} however, in our analysis the \textit{NuSTAR} spectrum does not require such an additional component (see Fig.~\ref{fig:spectra}). 
The addition of a Gaussian fixed at 6.4 keV indeed only leads to a $\Delta\chi^2 = 4.0$, which is clearly not significant. 
We also estimate that the equivalent width of the line is smaller than 0.03 keV. 
Therefore, the line may be present but our data are not of sufficient quality to clearly detect it. \par
The model that better reproduces the spectrum is a broken power law ($\chi^2_\nu = 1.11$ with 316 d.o.f., see Table \ref{tab:J0324}). 
This result is comparable to what is obtained from the physical model of thermal Comptonization, which seems to identify a soft excess below 3 keV. 
A single power law instead is not enough to reproduce the data. 
An F-test indicates that the probability that the fit improvement with a broken power law model was by chance is 5$\times$10$^{-3}$. 
A cut-off power law is not a good model for our data, while both a warm absorber and a black body seem to reproduce the spectrum reasonably well, although the broken power law remains the best representation of the data (see Table~\ref{tab:results}). \par
The broken power law model suggests the presence of a possible break around 13.40 keV, although the errors are fairly large and the break position is not very well constrained.
However, it is worth noting that the CompTT model and the broken power law represent the same physical scenario, since as mentioned before the latter is basically a phenomenological version of the former. The CompTT provides an estimate to the seed photon temperature, around T$_0 \sim$30$^{+12}_{-10}$ eV, which is comparable to what was found in other NLS1s both jetted and nonjetted (T$_0 \sim 30$ eV, \citealp{Larsson18, Chaudhury18}), or assumed as input parameter for the model in some other cases (10 eV, \citealp{Turner18, Younes19}). 
The sole CompTT model accounts in the 0.3-10 keV spectral region for a deabsorbed flux of 3.9$\times$10$^{-12}$ erg s$^{-1}$ cm$^{-2}$, which constitutes 25\% of the total flux of this spectral region, suggesting that a non-negligible thermal Comptonization component is present at low energies. 
A similar result is suggested by the black body model, which also identifies an additional component present at low energies. 
It is also possible that this component is due to the presence of a warm absorber, but these scenarios cannot be disentangled in our data. \par
The photon index measured below the break is close to that of unsaturated Comptonization ($\sim$1.9). 
Above the energy break the spectrum is significantly harder than that of typical nonjetted NLS1s (2.5, \citealp{Leighly99b}), as found already by \citet{Panessa11}, suggesting a jet contribution. 
However, both photon indexes above and belowe the break are softer with respect to the other sources in our sample, and also to objects in the literature \citep[e.g.,][]{Larsson18}. 
This is similar to what was observed during the high state of J0324, who noted that the spectrum during low states can be reproduced by a single soft power law, while in high states the jet component emerges, producing a spectral break and a hardening at high energies \citep{Foschini09, Foschini12}. \par

\begin{table}
\caption{Spectral fitting parameters for J1505.}
\centering
\footnotesize
\begin{tabular}{l l}
\hline
\multicolumn{2}{l}{\textbf{TBabs*zpo}} \\
\hline\hline
nH & 0.39$\times10^{21}$ \\
$\Gamma$ & 1.16$\pm$0.08 \\
norm$_1$ & (8.57$^{+1.38}_{-1.28}$)$\times$10$^{-5}$ \\
c$_{FPMA}$ & 0.55$\pm$0.10 \\
c$_{FPMB}$ & 0.52$\pm$0.10 \\
$\chi^2_\nu$/dof &  1.26/68 \\
Flux$^o_{0.3-70}$ & 3.89$\pm$0.62 \\
logL$^i_{0.3-70}$ & 45.19$\pm$0.16 \\
\hline\hline
\end{tabular}
\tablefoot{Lines: (1) Galactic hydrogen column density from \citet{Kalberla05} (cm$^{-2}$); (2) photon index of the redshifted power law; (3) normalization of the power law component; (4),(5) cross-calibration constants for FPMA and FPMB; (6) $\chi^2_\nu$ and degrees of freedom; (7) observed flux between 0.3 and 70 keV in units of $10^{-12}$ erg s$^{-1}$ cm$^{-2}$; (8) logarithm of the intrinsic luminosity between 0.3 and 70 keV.}
\label{tab:J1505}
\end{table}

\subsection{J0948}

The spectrum of J0948 is typically characterized by a soft excess on top of a hard power law \citep{Bhattacharyya14}.
Our simultaneous \textit{XMM-Newton}+\textit{NuSTAR} spectrum shown in Fig.~\ref{fig:spectra} confirms this result. 
Both the cut-off power law and the single power law can be immediately ruled out as representations of the spectrum (see Tables \ref{tab:J0948} and \ref{tab:results}). 
The broken power law provides a good fit for the data ($\chi^2_\nu = 1.06$, d.o.f. 1220), similarly to J0324. 
A spectral break is definitely present at 1.79$\pm$0.13 keV (F-test probability $<10^{-100}$), with photon indexes of 1.98$\pm$0.03 and 1.41$\pm$0.02 below and above the break, respectively. 
From a physical point of view the best result is obtained with a thermal Comptonization model on top of a power law component ($\chi^2_\nu = 1.01$, d.o.f. 1218). 
Our findings are consistent with an optically thick corona (although an optically thin corona cannot be ruled out given that the model provides only an upper limit for $\tau$; see Table \ref{tab:J0948}), with a seed photon temperature $\leq$81 eV, a value consistent with the corona of J0324.
In the range 0.3-10 keV, the flux we measured for the CompTT model only is 6.2$\times10^{-13}$ erg s$^{-1}$ cm$^{-2}$, which represents approximately 25\% of the flux in this spectral region. 
These coronal parameters are comparable to those derived for bright local Seyfert galaxies (both jetted and nonjetted, \citealp{Lubinski16, Tortosa18}), indicating that the soft excess of J0948 is similar to that of other AGN. 
The underlying power law has a rather hard photon index of 1.31$\pm$0.04, which is consistent with previous measurements for this source \citep{Foschini15, Paliya19}. 
As in the case for J0324, the X-ray spectrum of J0948 is likely a combination of a typical Seyfert emission and relativistic jet, with the latter being extremely prominent above a spectral break around 2 keV. 
Finally, we tried to model this source by adding a Fe K$\alpha$ line at 6.4 keV on top of the broken power law, but this feature is not significant ($\Delta\chi^2 =$3.9). 
If present, the line is extremely weak, with an equivalent width lower than 17 eV.
A warm absorber does not seem to be a good representation of the data, and the same is true for a simple black body. 
The latter partially reproduces the soft excess of J0948, but not as well as a Comptonization model.

\subsection{J1505}
This source was observed simultaneously with \textit{NuSTAR} and \textit{Swift/XRT} on 2017-02-12. 
However, the \textit{XRT} observation did not provide reasonably good coverage of the soft spectral region, and therefore we decided to add together ten \textit{XRT} observations in order to improve the statistics by one order of magnitude and to also model the part of the spectrum not covered by \textit{NuSTAR} (below 3 keV).
The downside of this procedure is that the addition of nonsimultaneous observations introduces some problems due to the source flux variability. 
For this reason, in the spectrum of Fig.~\ref{fig:spectra} the \textit{XRT} and \textit{NuSTAR} spectra are not fully consistent with each other. 
However, as mentioned above, the photon index remains stable during all observations.\par
The spectral modeling shows that the best fit is provided by a single power law (see Table \ref{tab:J1505}). 
Although the $\chi^2_\nu$ for a broken power law is lower, the F-test shows that the probability of this improvement occurring by chance is 0.12. 
This fairly high value suggests that the data are not of sufficient quality to safely claim that a broken power law is better than a single power law at describing the data.
The CompTT model is also not a good representation of the data. \par
The photon index provided by the single power-law model is 1.16$\pm$0.08, suggesting that the emission is strongly dominated by the relativistic jet. 
Such a photon index is rather hard, but it is in agreement with what was already found in the literature \citep{Foschini15, Paliya19}, especially during the high-activity state of the source \citep{Dammando16}. 
As previously mentioned, it is not obvious whether or not a Seyfert component is present. 
J1505 has the lowest flux in the sample, therefore there are only sparse data points in the spectrum that do not allow us to confirm the presence of a spectral break. 
However, past observations revealed that a broken power law may be present after all, suggesting that some coronal emission, although weak, could contribute below 2 keV \citep{Foschini15, Dammando16}. 

\begin{table}
\caption{Spectral fitting parameters for J2007.}
\centering
\footnotesize
\begin{tabular}{l l}
\hline
\multicolumn{2}{l}{\textbf{TBabs*(zpo+compTT)}} \\
\hline\hline
nH & 0.32$\times10^{21}$ \\
$\Gamma$ & 1.60$^{+0.03}_{-0.05}$ \\
norm$_1$ & (1.83$^{+0.05}_{-0.33}$)$\times 10^{-4}$ \\
T$_0$ & 9.9$^{+39.5}_{-9.0}$ \\
kT & 21.5$^{+54.2}_{-16.1}$ \\
$\tau_p$ & $\leq$75.1 \\
norm$_2$ & $\leq$112.5 \\
c$_{MOS1}$ & 1.06$\pm$0.04 \\
c$_{MOS2}$ & 1.01$\pm$0.04 \\
c$_{FPMA}$ & 1.38$\pm$0.10 \\
c$_{FPMB}$ & 1.39$\pm$0.11 \\
c$_{FPMA}$ & 1.62$\pm$0.10 \\
c$_{FPMB}$ & 1.49$\pm$0.10 \\
$\chi^2_\nu$/dof &  0.99/373 \\
Flux$^o_{0.3-70}$ & 2.56$\pm$0.26 \\
logL$^i_{0.3-70}$ &  44.50$\pm$0.10 \\
\hline\hline
\end{tabular}
\tablefoot{Lines: (1) Galactic hydrogen column density from \citet{Kalberla05} (cm$^{-2}$); (2) photon index of the redshifted power law; (3) normalization of the power law component; (4) seed photon temperature (eV); (5) plasma temperature (keV); (6) optical depth of the plasma; (7) normalization of thermal Comptonization model; (8),(9),(10),(11) cross-calibration constants for FPMA and FPMB; (10) $\chi^2_\nu$ and degrees of freedom; (12) observed flux between 0.3 and 70 keV in units of $10^{-12}$ erg s$^{-1}$ cm$^{-2}$; (13) logarithm of the intrinsic luminosity between 0.3 and 70 keV.}
\label{tab:J2007}
\end{table}

\subsection{J2007}
This source has been observed twice with \textit{NuSTAR} (Table \ref{tab:obs_nustar}), and an \textit{XMM-Newton} observation is quasi-simultaneous to first of the \textit{NuSTAR} observations (see Table \ref{tab:obs_xmm}). 
Considering that the flux levels of the source did not change significantly between \textit{NuSTAR} observations and that the photon indexes are consistent in the common energy range, we decided to analyze both spectra simultaneously along with the \textit{XMM-Newton} data. \par
The various models we used are all good representations of the observed data in terms of $\chi^2_\nu$, with the poorest being the one including an ionized Fe line and the warm absorber which are both likely overfitting the data.
A broken power law in this case may be slightly better than a single power law at reproducing the data. 
The F-test suggests indeed that there is only a 1\% probability that the $\chi^2$ improvement is due to chance. 
The spectral break is found at 2.01 keV, while the photon indexes are 1.66 and 1.56 below and above the break, respectively (see Table~\ref{tab:results}). 
The CompTT model also suggests that a very small thermal Comptonization component is present below 1 keV, and the same result is confirmed by the black body model.  
The seed photon temperature is $\sim$10 eV and the electron temperature of $\sim$20 keV, possibly suggesting a slightly warmer corona than in J0324 and J0948. 
It is nevertheless worth noting that the flux estimated by the sole CompTT model between 0.3 and 10 keV is 1.76$\times10^{-14}$ erg s$^{-1}$ cm$^{-2}$, which accounts for $\sim$2\% of the total flux in this spectral region. 
After 1 keV, the CompTT model contribution is completely negligible, and only a possible jet component remains present. 
The thermal Comptonization component is therefore weak when compared to those of other objects. 
This result is consistent with what was found during the \textit{XMM-Newton} observation in 2004 \citep{Gallo06} but not confirmed by subsequent observations \citep{Kreikenbohm16}. 
Given the very small contribution brought from the corona to the spectrum, a lower signal-to-noise ratio or an increased jet emission can both render it invisible. \par

\section{Time variability in NuSTAR}

\begin{table*}
\caption{Parameters derived from the light curves.}
\label{tab:fit_parameters}
\centering
\footnotesize
\begin{tabular}{l c c c c c c c}
\hline\hline
Source & Model & F$_{const}$ & t$_p$ & F$_{peak}$ & $\tau_r$ & $\tau_d$ & $\chi^2_\nu$ \\
\hline
J0324 & C & 0.56$\pm$0.07 & {} & {} & {} & {} & 14.5 \\
J0324 & F & 0.54$\pm$0.01 & 96.1$\pm$1.3 & 0.33$\pm$0.04 & 4.6$\pm$1.0 & 7.7$\pm$1.0 & 7.4 \\
J0948 & C & 0.13$\pm$0.01 & {} & {} & {} & {} & 1.9 \\
J1505 & C & (4.0$\pm$0.05)$\times$10$^{-2}$ & {} & {} & {} & {} & 0.8 \\
J2007 A & C & (7.3$\pm$1.0)$\times$10$^{-2}$ & {} & {} & {} & {} & 1.6 \\
J2007 A & F & (6.7$\pm$0.2)$\times$10$^{-2}$ & 59.3$\pm$2.4 & $\leq$0.32 & $\leq$3.9 & 8.5$\pm$7.6 & 0.9 \\
J2007 B & C & (7.8$\pm$1.5)$\times$10$^{-2}$ & {} & {} & {} & {} & 4.7 \\
J2007 B & F & (6.5$\pm$0.3)$\times$10$^{-2}$ & 62.2$\pm$2.7 & (4.7$\pm$0.8)$\times$10$^{-2}$ & 4.2$\pm$2.8 & 24.0$\pm$9.3 & 1.4 \\
\hline\hline
\end{tabular}
\tablefoot{Columns: (1) source; (2) model used to represent data. C stands for constant flux, F stands for exponential flare; (3) flux continuum level (counts per second); (4) peak position (in ks after the observation start); (5) maximum flux above the continuum level (cps); (6) time scale of the rise in the source frame (ks); (7) time scale of the decay in the source frame (ks); (8) reduced chi-squared.}
\end{table*}

\begin{table}[!t]
\caption{Consecutive point variability parameters for J0324.}
\label{tab:0324_fast}
\centering
\footnotesize
\begin{tabular}{l c c c c }
\hline\hline
Time & $\tau_2$ & F$_0$ & signif. & R/D \\
\hline
22.35 & 17.25$\pm$4.54 & 0.61$\pm$0.02 & 6 & D \\
63.04 & 18.17$\pm$4.34 & 0.64$\pm$0.02 & 6 & D \\
132.81 & 18.05$\pm$4.18 & 0.49$\pm$0.02 & 7 & R \\
190.95 & 17.11$\pm$4.86 & 0.55$\pm$0.02 & 6 & D \\
\hline
\end{tabular}
\tablefoot{Columns: (1) time (ks) after the observation start; (2) doubling time (ks) in the source restframe; (3) initial flux (cps); (4) significance of the variation (in $\sigma$ units); (5) rise (R) or decay (D) of the light curve.}
\end{table}

We extracted the light curves of each source between 4 and 50 keV during the \textit{NuSTAR} observation by co-adding the data of FPMA and FPMB. 
We reproduced each one of them by fitting it with a constant flux.
When the $\chi^2_\nu$ indicated that there is a significant deviation from this behavior, and an increased activity is clearly evident from the light curve, we tried to add an exponential flare component to the constant flux, assuming as initial guess that the maximum flux reached during the observation corresponds to the flare peak. 
The flare was modeled with an asymmetric exponential rise and decrease of the flux \citep{Valtaoja99}, in the form
\begin{equation}
\Delta F(t) = \begin{cases}  \Delta F_{max} e^{(t - t_{peak})/\tau_r}\\\Delta F_{max} e^{(t_{peak} - t)/\tau_d} \end{cases} \; ,
\label{eq:exp}
\end{equation}
where $F$ is the source flux, t$_{peak}$ is the time where the peak flux F$_{max}$ is reached, $\tau_r$ is the timescale of the rise, and $\tau_d$ is the timescale of the decay. 
The parameters and their associated errors were derived with a Monte Carlo method, by adding Gaussian noise to each flux measurement proportional to its error bar, repeating the fit with the different flux values 100 times, and calculating the median value and its standard deviation for each parameter. 
Finally, when the $\chi^2_\nu$ indicated some residual variability, we followed the approach of \citet{Foschini11a}. 
We checked consecutive points of the light curves, looking for variability at a significance level of at least 5$\sigma$. 
For each of these episodes we estimated the doubling timescale following the relation
\begin{equation}
F(t_2) = F(t_1) \cdot 2^{-(t_2 - t_1)/\tau_2}
\end{equation}
where F(t$_2$) and F(t$_1$) are the fluxes of two consecutive points at t$_1$ and t$_2$, and $\tau_2$ is the doubling timescale. 
The errors were estimated using a Monte Carlo method. 
We added a Gaussian noise to the flux of each of the two consecutive points proportional to their respective errors, and repeated the measurements 100 times.

\subsection{J0324}

\begin{figure}
\includegraphics[width=\hsize]{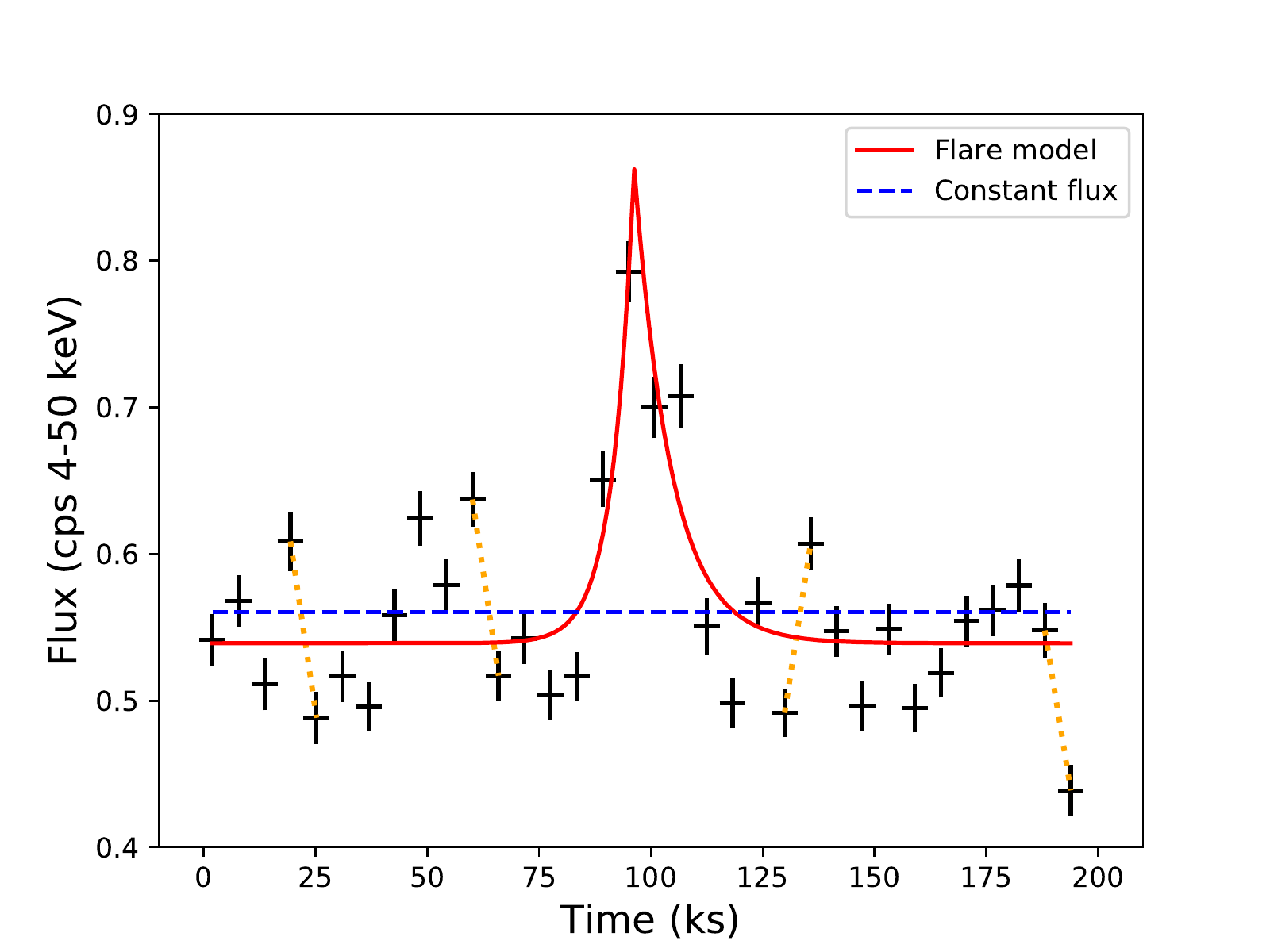}
\caption{Light curve between 4 and 50 keV of J0324 during the \textit{NuSTAR} observations. The blue dashed line indicates the constant flux model. The red solid line represents the asymmetric exponential flare model that we applied to the curve. The dotted orange lines represent the consecutive points where the flux changes by more than 5$\sigma$.}
\label{fig:flare_0323}
\end{figure}

In the light curve we detect a possible small flare of the source, analogous with what was found by \citet{Landt17}. 
Our findings indicate an improvement of $\Delta\chi^2_\nu = 7.1$ when the exponential flare model is added to a constant flux. 
The light curve and best fit performed using Equation \ref{eq:exp} are shown in Fig.~\ref{fig:flare_0323}. 
The parameters are shown in Table \ref{tab:fit_parameters}. 
According to our fit, the peak is reached 96.1 ks after the observation start, when the flux reaches 0.33 counts s$^{-1}$ above the continuum level, corresponding to an approximately 38\% increase in the flux. 
The timescale of the flare, corrected for redshift, is 4.6 ks during the rise, and 7.7 ks for the decrease.  
Even with the flare model, the $\chi^2_\nu$ is very high. 
Indeed there is a strong residual variability that must be accounted for with variability between consecutive points. 
As shown in Table \ref{tab:0324_fast}, we found four cases in addition to the previously modeled flare in which the flux difference between consecutive points is larger than 5$\sigma$, all with a similar doubling time of $\sim$18 ks. \par
We also analyzed the light curve in soft and hard X-ray band, and estimated the hardness ratio following the usual
\begin{equation}
HR = \frac{H - S}{H + S} \; ,
\label{eq:hr} 
\end{equation}
where H is the flux between 10 and 50 keV, and S is the flux between 4 and 10 keV. 
The hardness ratio is indeed consistent with a constant for the whole observation, with a mean value of $-0.37$ and a standard deviation from this value of 0.02.  
The constant value for the flare indicates that the flux variations during flaring and fast variability episodes occur simultaneously throughout the entire spectrum. 
As shown in Fig.\ref{fig:spectra}, above 4 keV the jet power law dominates over the Seyfert component. 
Therefore, this flare is likely associated with relativistic jet. \par
The BH mass of jetted NLS1s is typically in the range of 10$^7$-10$^8$ M$_\odot$ \citep{Foschini15, Berton15a}.  
For J0324, the most accurate estimate using reverberation mapping is 3.4$\times10^7$ M$_\odot$ \citep{Wang16}, which in terms of gravitational radius corresponds to r$_g \sim$5$\times$10$^{12}$ cm. 
The minimum variability time corresponding to the light travel time of the BH horizon is then r$_g$/c $\sim$ 3 minutes, significantly shorter than all the variability timescales we measured. 
The size of the plasma blob derived from the minimum timescale $\tau$ (1.3 hr) observed during flares is r $<$ c$\delta\tau \sim 300$ r$_g$, by assuming a typical Doppler factor $\delta \sim 10$ \citep{Ghisellini10}. 
The blob size is directly connected to its distance $R$ from the central engine, following r $\sim \phi R$, where $\phi$ is the semi aperture of the jet cone. 
For a typical blazar, the opening angle is 1.3$^\circ$ \citep{Pushkarev17}, and therefore the dissipation region scale is lower than 6$\times10^{16}$ cm, corresponding to $R \sim$(10$^4$)$\times$ r$_g$. 
This scale is similar to that of the broad-line region (BLR, 4$\times$10$^{16}$ cm for J0324, \citealp{Foschini15}), suggesting that the dissipation occurred inside the BLR, as observed in many FSRQs \citep{Ghisellini10}.

\subsection{J0948 and J1505}
\begin{figure}[!t]
\includegraphics[width=\hsize]{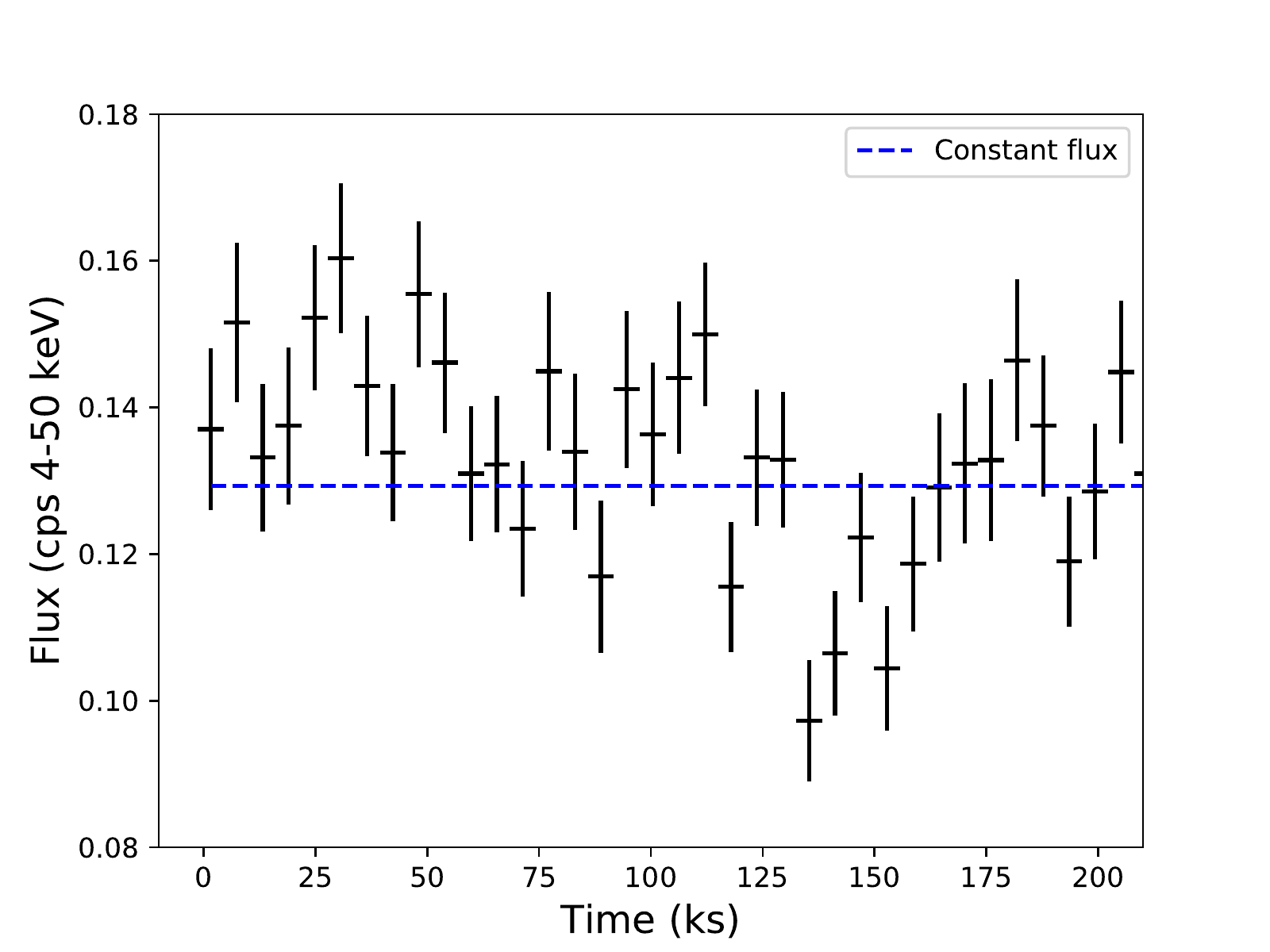}
\caption{\textit{NuSTAR} light curve between 4 and 50 keV of J0948. The blue dashed line indicates the constant flux model. }
\label{fig:flare_0948}
\includegraphics[width=\hsize]{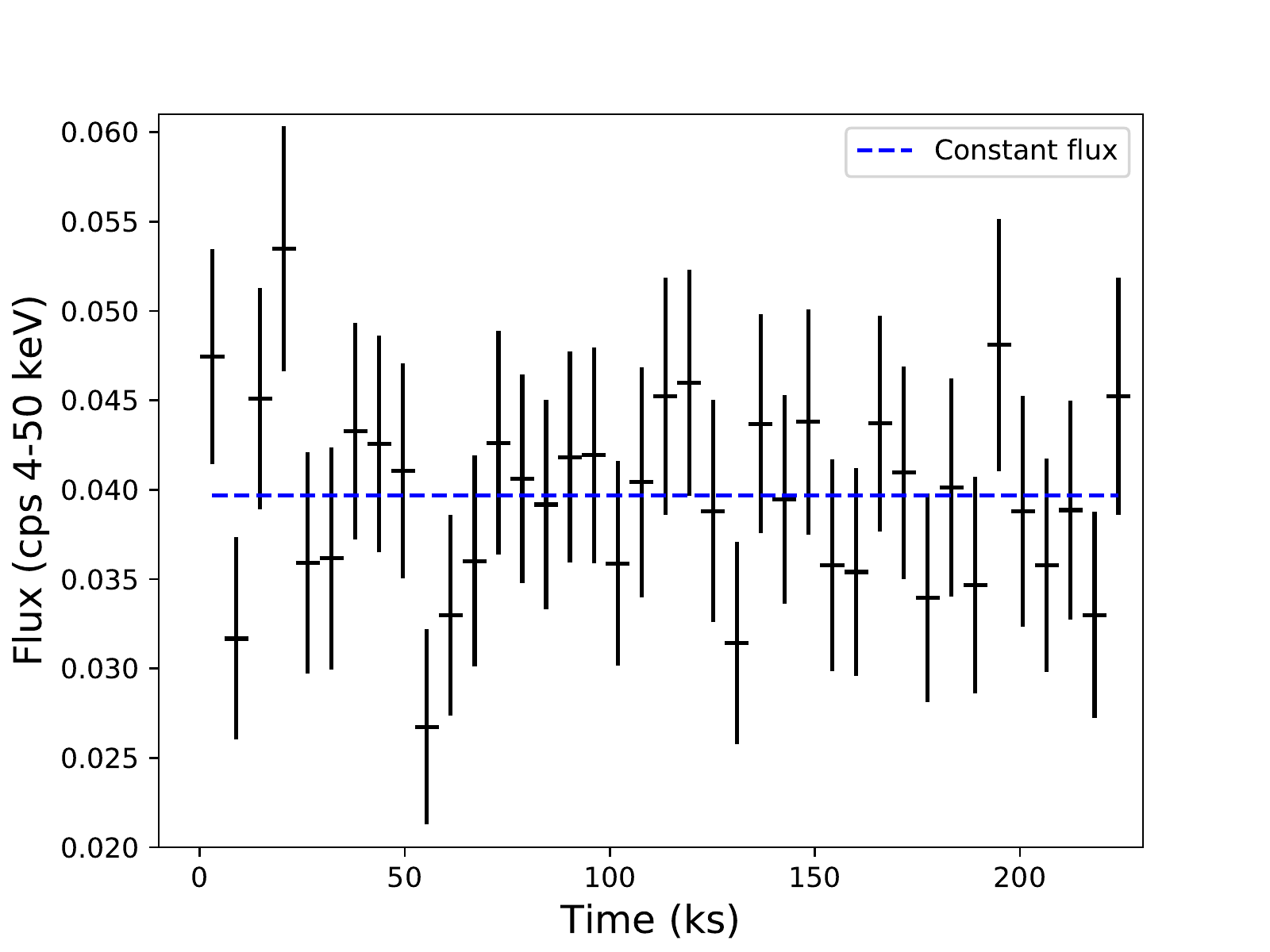}
\caption{\textit{NuSTAR} light curve between 4 and 50 keV of J1505. The blue dashed line indicates the constant flux model. }
\label{fig:flare_1505}
\end{figure}

For both J0948 (Fig.~\ref{fig:flare_0948}) and J1505 (Fig.~\ref{fig:flare_1505}) the light curves are not characterized by any significant variability. 
In both sources, the $\chi^2_\nu$ analysis indicates that the constant flux model is enough to reproduce the data (see Table \ref{tab:fit_parameters}).
Some minor variability is present in J0948, but no consecutive points have a flux variation larger than 5$\sigma$. 
The same is true for J1505, even though in this case the variability is even lower. 
The hardness ratio, as in the rest of the light curve, does not show any significant variability. 
For J0948 we measured a HR of $-$0.15 and a standard deviation of 0.07, while for J1505 the mean HR is $+$0.01 with a standard deviation of 0.14. 
These values are consistent with what was found from the spectra, with the highest HR found in J1505, as expected from the source with the hardest photon index of our sample.

\begin{figure}[!t]
\includegraphics[width=\hsize]{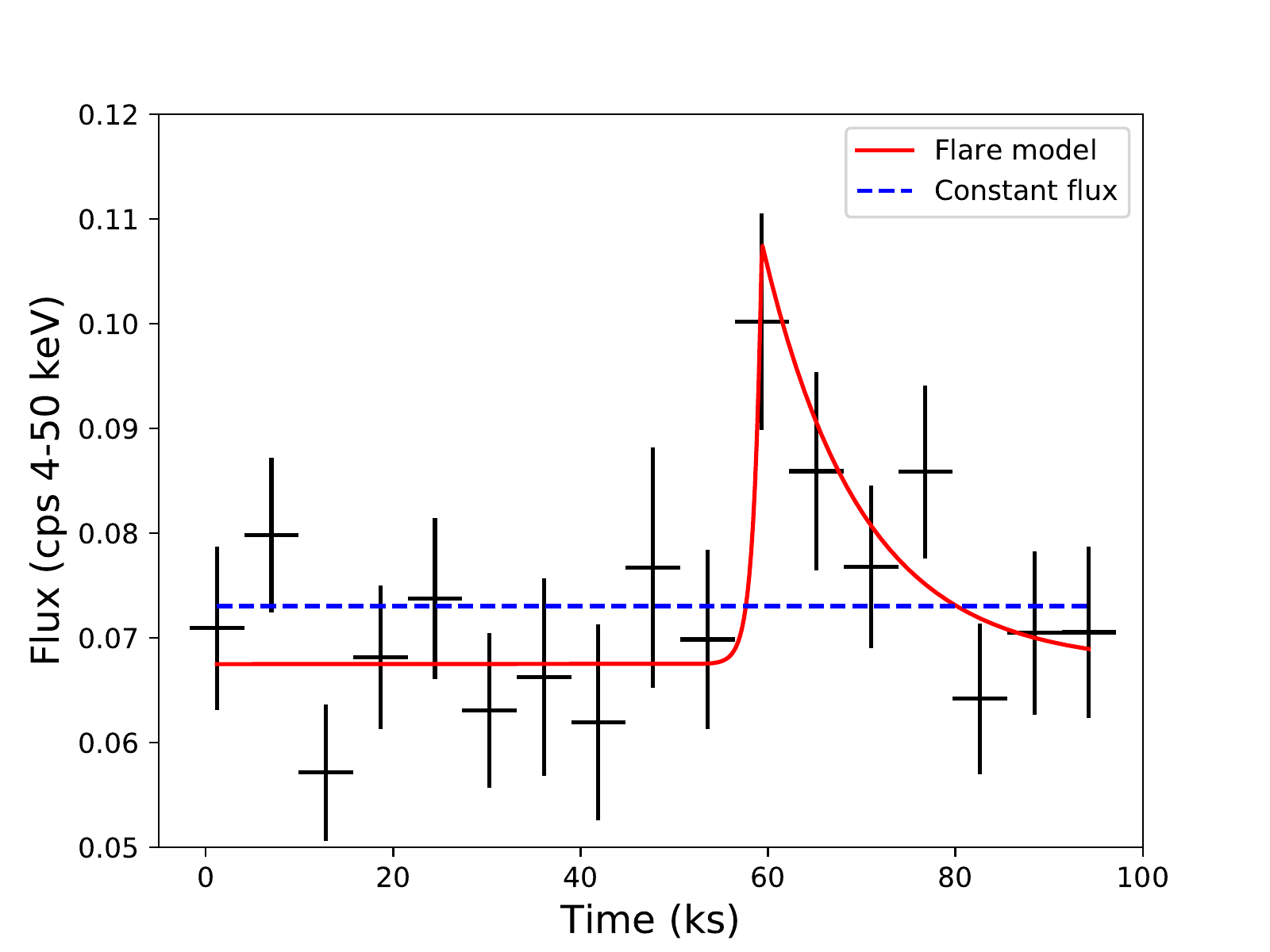}
\caption{Light curve between 4 and 50 keV of J2007 during the \textit{NuSTAR} observation of 2016-05-09. The blue dashed line indicates the constant flux model. The red solid line represents the asymmetric exponential flare model that we applied to the curve. }
\label{fig:flare_2004A}
\includegraphics[width=\hsize]{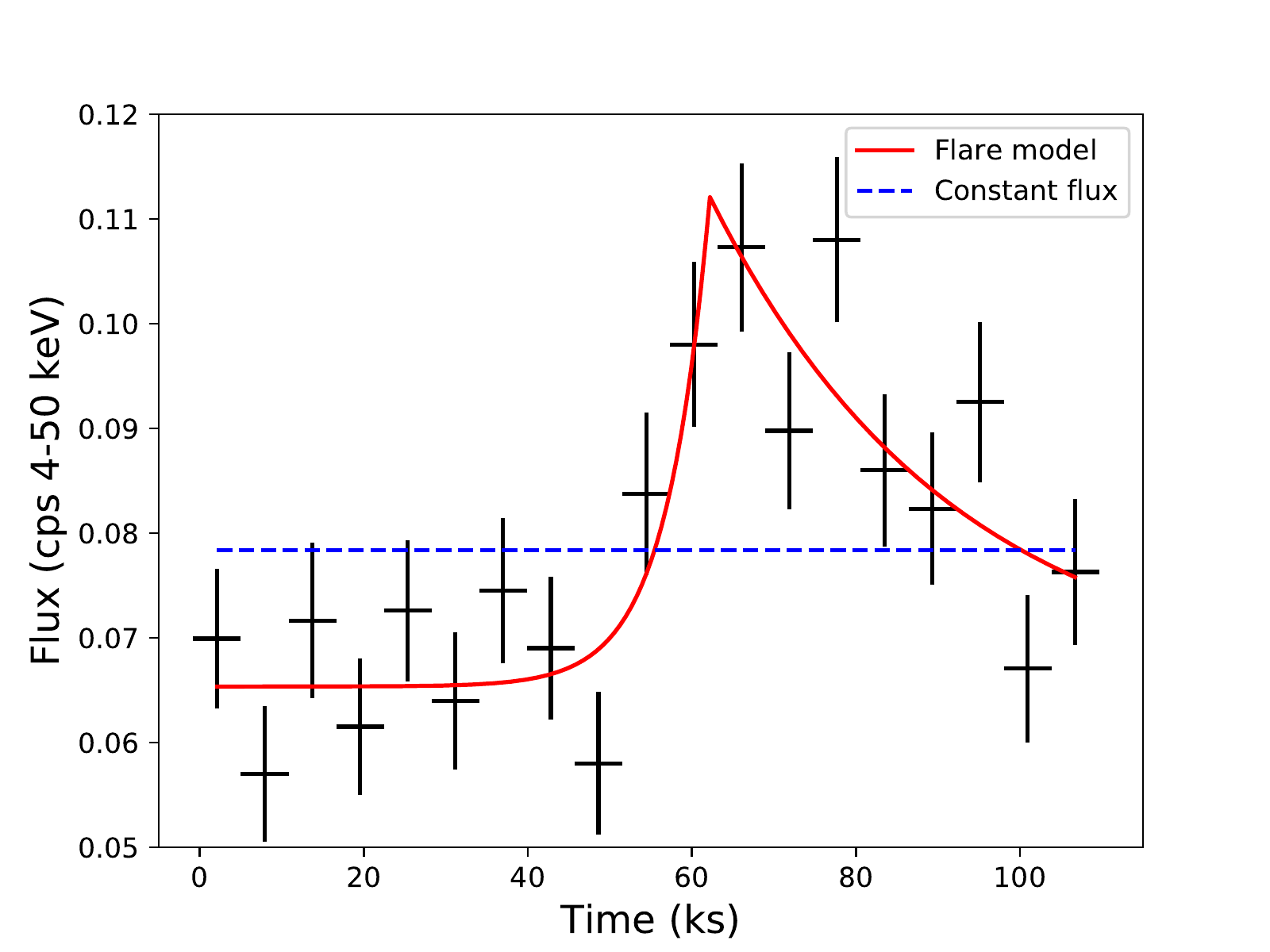}
\caption{Light curve between 4 and 50 keV of J0324 during the \textit{NuSTAR} observation of 2016-10-23. The blue dashed line indicates the constant flux model. The red solid line represents the asymmetric exponential flare model that we applied to the curve. }
\label{fig:flare_2004B}
\end{figure}

\subsection{J2007}
\label{sec:flare_J2004}
In J2007, the first light curve (2016-05-09) is well reproduced by a constant flux, while the second (2016-10-23) is not (see Table \ref{tab:fit_parameters}). 
However, upon visual inspection of the first light curve we see a sudden increase of the flux followed by a possible decay, we tried to add an asymmetric exponential flare model in both of data sets. 
The $\chi^2_\nu$ decreases by 0.7 in the first observation of 2016-05-09, and by 3.3 in the second. 
For the first observation (Fig. \ref{fig:flare_2004A}), after the addition of the flare model, the $\chi^2_\nu$ may indicate an overfitting of the data, so it could still be possible that this feature is not really present. 
In the second observation, shown in Fig.~\ref{fig:flare_2004B}, there is a $\sim$40\% flux increase with respect to the constant level, and the decay timescale is significantly larger than the rise time. 
However, errors are rather large, since this second flare occurred fairly close to the end of the observation, and the decay timescale is not very well constrained.
In both light curves the consecutive point variability approach was not used because we found no flux changes above 5$\sigma$. 
In terms of hardness ratio, analogously with what was found in J0324, no variability seems to be present, with a mean value of $-$0.24 and standard deviation 0.09 for the first observation, and a mean of $-$0.23 and standard deviation of 0.11 in the second one. 
This confirms, as expected, that the origin of this variability resides in the relativistic jet. \par
As for J0324, we tried to estimate the physical parameters of the emitting region. 
The BH mass of J2007 is 7.0$\times$10$^7$ M$_\odot$ \citep{Foschini15}, which corresponds to a r$_g \sim 9\times 10^{12}$ cm and a minimum variability time of 340 s. 
The shortest timescale observed for this source is the upper limit found for the light curve of 2016-05-09, that is 3.9 ks\footnote{Although this value is derived from a low-confidence flare, the result does not significantly change when the shortest time scale of 2016-10-23 (4.2 ks) is used.}.
Using the same assumptions as above, this indicates an emitting region radius size of r $<$ 1$\times$10$^{15}$ cm, which for a typical semi-aperture of the jet cone for blazars corresponds to 4$\times 10^{16}$ cm. 
Expressed in gravitational radii, the dissipation region scale is 4.3$\times 10^{4}$ r$_g$, which is smaller than the scale of the molecular torus \citep[e.g.,][]{Kynoch18}.
Therefore, also in this case the flare probably occurred within the BLR. 
However, it is worth noting that in the same source it is possible that different flares originate in different regions \citep{Foschini11a}. 
Indeed, during outbursts high-energy photons can be produced farther from the nucleus, close to the molecular torus \citep[e.g.,][]{Donea03, Aleksic11a, Aleksic11b, Tavecchio13, Coogan16}. 
A better constraint on the position of the emission region may come from future observations with instruments such as the Cherenkov Telescope Array (CTA, \citealp{Romano18}, Romano et al., in prep.). 

\section{$\gamma$-NLS1s on the fundamental plane}

\begin{figure*}
\includegraphics[width=0.5\hsize]{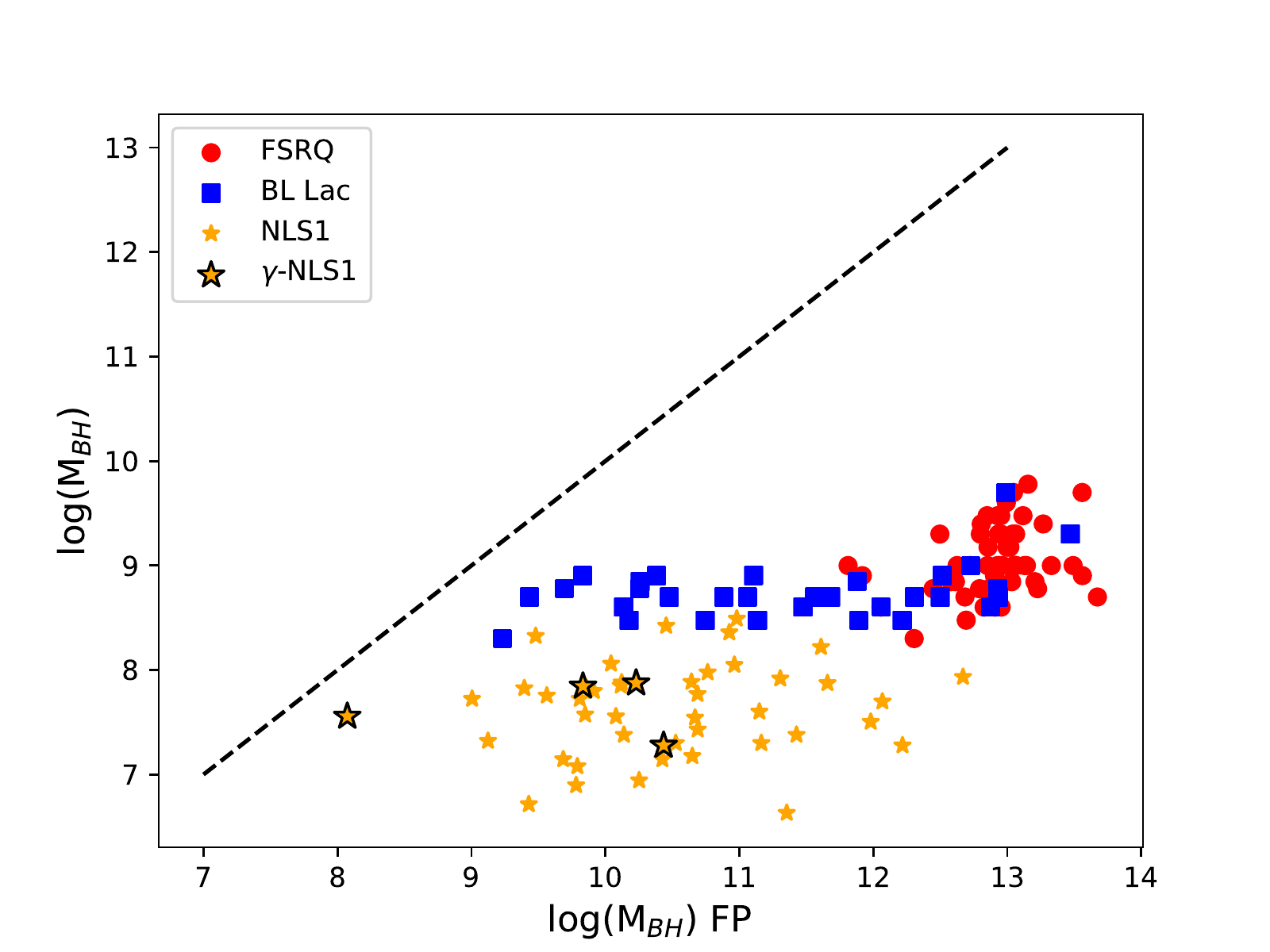}
\includegraphics[width=0.5\hsize]{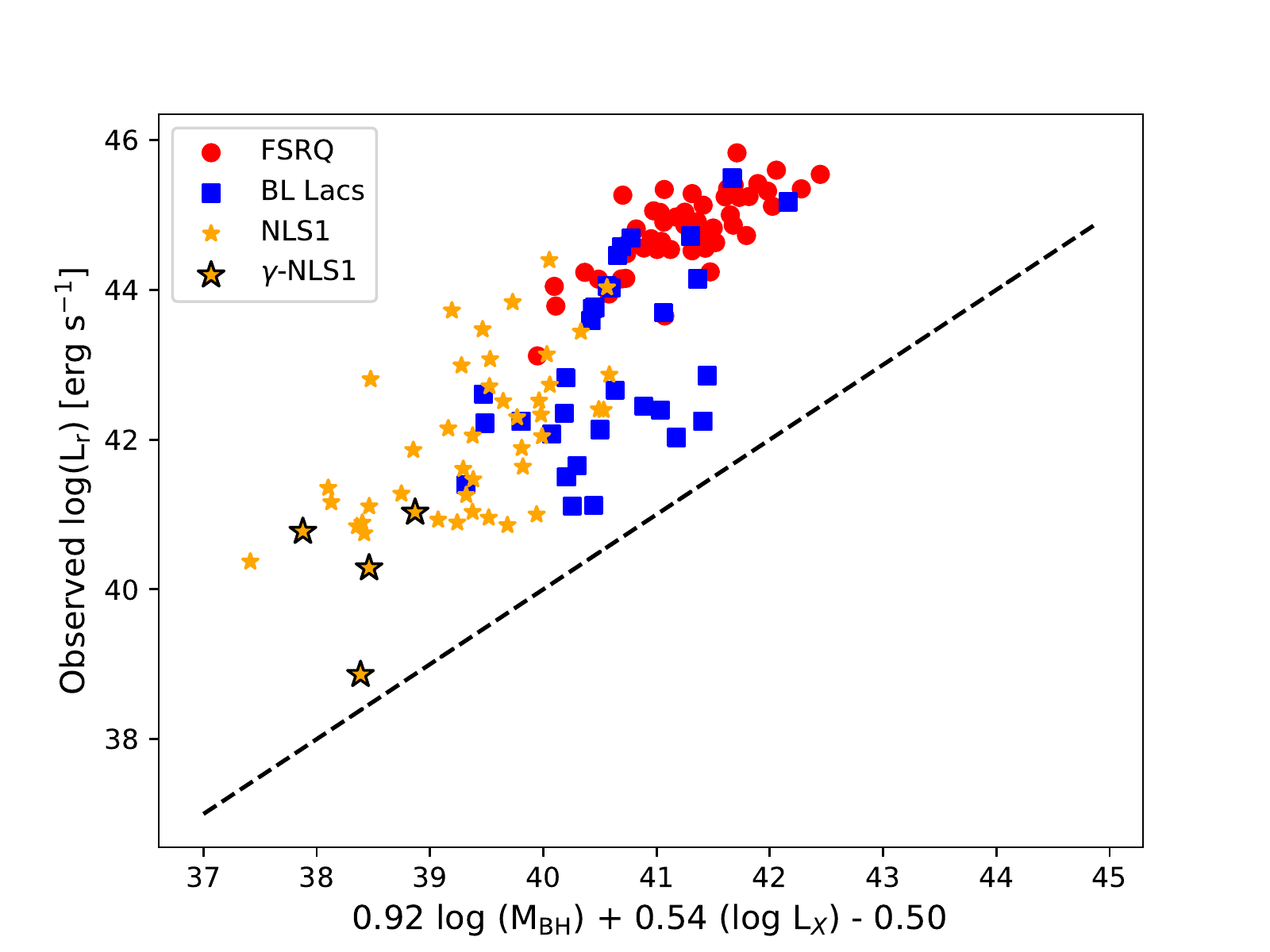}
\caption{\textbf{Left:} Black hole masses calculated by means of the fundamental plane as reported by \citet{Gultekin19}, and masses estimated via optical spectroscopy. \textbf{Right:} Radio luminosity as estimated from the fundamental plane, and measured radio luminosity. Red circles represent FSRQs, blue squares represent BL Lacs, and orange stars indicate NLS1s. The larger stars with a black edge are the four sources we analyzed. The black dashed line is the bisector. } 
\label{fig:fp} 
\end{figure*}

An interesting application of the X-ray luminosity is the so-called fundamental plane (FP) of BH activity. 
For many years it has been known that both AGN and Galactic BHs follow this empirical relation \citep{Merloni03}, which is a correlation between radio luminosity at 5 GHz, X-ray luminosity in the range 2-10 keV, and BH mass. 
The physics underlying the FP is the nonlinear scalability between accretion, jet power, and BH mass \citep{Heinz03}, which is fundamental in the unification of relativistic jets \citep{Foschini14}. 
When projected, the FP can be expressed as 
\begin{equation}
\log \left( \frac{M}{M_\odot} \right) = a + b \log \frac{L_{5 \; \rm{GHz}}}{\rm{erg \; s}^{-1}} + c \log \frac{L_{2-10 \; \rm{keV}}}{\rm{erg \; s}^{-1}} \; .
\label{eq:fp}
\end{equation} 
This relation could be used in principle as an independent indicator of BH mass, which is a crucial parameter to understand the nature of NLS1s. 
Some authors indeed believe that the narrowness of permitted lines is not connected to the BH mass, but that it must be explained by means of the inclination effect. 
If the BLR had a flattened geometry when observed pole-on, as observed in other AGN \citep{Gravity18}, no rotation component would be directed along the line of sight. 
This would produce a lack of Doppler broadening, which could account for the narrow lines \citep{Decarli08}. 
In the case of $\gamma$-ray-emitting NLS1s, this problem could be particularly severe, since the presence of $\gamma$-ray emission already indicates that they are observed pole-on. \par
We tried to estimate the BH mass using the coefficients of Equation \ref{eq:fp} derived by \citet{Gultekin19} from a sample of mostly nonjetted AGN ($a = 0.55\pm0.22$, $b = 1.09\pm0.10$, and $c = -0.50^{+0.16}_{-0.15}$). 
For the X-ray luminosity, we used the intrinsic luminosity derived from our data (see Table \ref{tab:fluxes}).
For J0324, J0948, and J2007 we adopted the Comptonization model, while for J1505 we used the single power law model. 
The intrinsic luminosity was estimated from the flux by correcting for Galactic absorption and applying a K-correction based on our models. 
For the radio flux at 5 GHz, we used values retrieved in the literature and shown along with their references in Table \ref{tab:source_list}. \par
For J0324 we derived a BH mass of 2$\times$10$^9$ M$_\odot$, and for all the other sources the masses are all close or above 10$^{11}$ M$_\odot$. 
These values are clearly not physical. 
The maximum mass for an accreting BH was predicted to be of the order of 10$^{10}$ M$_\odot$ \citep{Inayoshi16}, and for J0324 this value is extremely far from all the other estimates in the literature \citep[e.g., see][and references therein]{Landt17}. \par
However, the effects of relativistic beaming have not been considered in our calculations. 
The correcting factor $f$ to apply would be
\begin{equation}
f = 1 - \cos\frac{1}{\Gamma_b} \;
\end{equation}
where $\Gamma_b$ is the bulk Lorentz factor of the jet. 
For our sources, this factor is approximately ten \citep{Abdo09c}.
Both radio and X-ray luminosity should be corrected as $L_{unbeamed} = L_{obs} \times f$. 
Despite this, all the BH masses for our objects remain quite far from the values derived by \citet{Foschini15} or any other estimate in the literature \citep[e.g.,][]{Yuan08}. 
The deviations from the virial mass estimate go from 0.5 dex for J0324 (1$\times10^8$ M$_\odot$), to 2.3 dex (1.1$\times10^{10}$ M$_\odot$) for J1505.
It is worth noting that the smallest discrepancy is observed in J0324, which among our sourcse has by far the lowest radio luminosity and the strongest accretion disk \citep{Foschini15}. 
Indeed, the main cause of these large deviations is the presence of the relativistic jet, which is dominant especially in the radio domain. 
The strongest dependency of the FP is indeed on radio luminosity, with a positive correlation with BH mass. 
X-rays instead provide an inverse contribution, with lower X-ray luminosities for higher BH masses. \par
This result can be found in other jetted AGN. 
We tried to use the FP relation on the blazars of the sample by \citet{Ghisellini10}, obtaining very similar results. 
The BH masses are all clearly overestimated, with a mean value of 8.1$\times$10$^{12}$ M$_\odot$ for FSRQs and a standard deviation of 0.4 dex, and 2.4$\times$10$^{11}$ M$_\odot$ with a standard deviation of 1.1 dex for BL Lacs. 
We also performed the same test on the sample of F-NLS1s by \citep{Foschini15}, and found an average BH mass of 1.9$\times$10$^9$ M$_\odot$, again well above the standard values of jetted NLS1s \citep[e.g.,][]{Komossa18}. \par
A strong discrepancy between the "classic" fundamental plane of BH accretion and blazars has already been found by \citet{Zhang18}. 
Our findings seem to confirm their result. 
Jetted AGN, especially powerful blazars, may indeed follow a different FP relation with respect to nonjetted sources. 
This inconsistency is immediately evident in the left panel of Fig.~\ref{fig:fp}, where the BH mass estimated via fundamental plane and via optical spectroscopy for our sources and the samples of NLS1s, FSRQs, and BL Lacs mentioned above are shown together. 
An identical result is shown in the right panel of Fig.~\ref{fig:fp}, in which the plane is represented in terms of radio luminosity, as it is shown in most of the literature. \par
The presence of the relativistic jets strongly affects the position of jetted NLS1s on the fundamental plane. 
This is because of relativistic beaming, the fast variability of jetted NLS1s, which typical of blazars \citep[e.g.,][for J0324 and J1505]{Ojha19}, and the different origin of X-ray and radio emission with respect to nonjetted sources. 
Variability in the jet activity state also affects the ratio between the corona and the jet flux, with the former being evident when the jet contribution is lower \citep{Grandi04, Foschini06}.
The radio emission, instead, is produced by the synchrotron emission of the jet, although a star formation component cannot be excluded \citep{Caccianiga15}.  
Conversely, in nonjetted NLS1s, the X-ray emission only originates in the corona, and the radio emission is mainly produced by star formation (although a nonthermal component from a jet base may also be present, see \citealp{Giroletti09}). 
In conclusion, the FP does not seem to be a good tool to estimate BH masses of $\gamma$-NLS1s and other blazars in general. 
It is interesting how in Fig. 5 of \citet{Merloni03} the blazar 3C 273 was the source showing the largest deviation from the plane relation. 
This could already be considered as a warning against the use of this tool for blazars. \par
In the NLS1 case, the use of the FP relation to estimate the BH mass can be difficult even for regular "radio-quiet" NLS1s. 
As shown by \citet{Lahteenmaki18}, among radio-quiet NLS1s there is a population of jetted objects which were never detected in radio surveys. 
A simple explanation for this is the nonlinear scalability of the jet power with the BH mass \citep{Heinz03}. 
In NLS1s, which are characterized by low-mass BHs, the jet power is lower than in other blazars, and this translates into lower radio (and $\gamma$ ray) luminosities. 
Such objects were simply invisible to all past surveys during their quiescent state and became apparent only during flares. 
However, the jet contribution to their X-ray and radio emission may still be significant, thus affecting any mass estimate via FP.

\section{NLS1s on the blazar sequence}
Above 10 keV, the spectra of J0948 and J1502 can be described with a power law, with rather hard photon indexes (1.30 and 1.16, respectively). 
This is a behavior typical of blazars, where the spectrum is dominated by the featureless continuum of the jet. 
However, an inverse Comptonization component is clearly present in J0948, and may also contribute to the spectra of J0324 and J2007.
This behavior is similar to what is observed in FSRQs, where a blazar component co-exists along with a Seyfert component \citep{Grandi04}. \par
According to the blazar sequence \citep{Fossati98, Ghisellini98, Ghisellini16}, the two classical blazar classes (FSRQs and BL Lacs) are characterized by the double-humped SED, with the first hump originating from synchrotron radiation, and the second one from IC. 
The blazar sequence is usually interpreted in terms of electron cooling and the surrounding environment. 
In FSRQs the environment is photon-rich, and the cooling process occurs easily via IC mechanisms, especially EC. 
In BL Lacs the situation is the opposite, and the main cooling mechanism is synchrotron self-Compton (SSC). 
In the \textit{NuSTAR} energy range the SED shape provides a rather clear division between the two classes. 
BL Lacs, in particular TeV sources, have a softer spectrum due to the tail of the synchrotron hump ($\Gamma >$ 1.5, \citealp{Costamante18}). 
Conversely, FSRQs have a much harder spectrum, associated with the rise of the IC hump \citep{Padovani02, Landt08, Bhatta17}. 
As confirmed by this analysis, $\gamma$-NLS1s have a rather hard spectrum, similar to that of FSRQs. \par
Despite this similarity in the SEDs, in terms of luminosity NLS1s are dimmer with respect to the most powerful blazars.
This behavior can be accounted for in the framework of the jet-BH mass scalability \citep{Heinz03, Foschini14}. 
If the jet power is proportional to the BH mass, the relatively low mass of NLS1s translates to a lower luminosity and a similar shape of the SED, which is what we indeed observe. 
In conclusion, our results seem to confirm that $\gamma$-NLS1s are essentially small-scale FSRQs \citep[e.g.,][]{Foschini15, Paliya16, Paliya16a, Berton16c, Berton17, Berton18a, Paliya19}, because the low BH mass can straightforwardly explain the lower jet power and luminosity of F-NLS1s with respect to FSRQs. \par
It is also worth noting that if F-NLS1s were only a projection effect due to a disk-like BLR viewed pole-on, there would be a contradiction in the physical mechanism accounting for the blazar sequence. 
If their jet power was low despite the presence of a large BH mass (like in BL Lacs), this would mean that there are relativistic electrons that are not cooling despite the photon-rich environment of NLS1s \citep{Foschini17}. 
Since this is not possible, we have to conclude that the difference in jet power and luminosity originates from the different BH mass. 
In conclusion, NLS1s do not fit the classical blazar sequence because their behavior does not depend only on the electron cooling. 
The sequence is therefore missing a key ingredient: the evolution of sources \citep[see Fig.~1A in][]{Foschini17}.

\section{Summary}
Here, we analyze the X-ray spectra observed with \textit{NuSTAR}, \textit{XMM-Newton} and \textit{Swift/XRT} of four bright $\gamma$-ray-emitting NLS1s, 1H 0323+342, PMN J0948+0022, PKS 1502+036, and PKS 2004-447. 
In all these sources, the spectrum above $\sim$2 keV is dominated by the nonthermal emission from their beamed relativistic jet. 
This is suggested by the fact that the photon indexes are much harder than those observed in nonjetted NLS1s, and are instead comparable to those observed in FSRQs. 
Below 2 keV, a soft excess possibly associated with a hot corona is present at different levels in three out of four sources. \par
We also examined the light curve of our sources between 4 and 50 keV. 
In J0324 and J2007 we detected short exponential flares, likely due to the fast variability of the jet. 
Using the timescales derived from the variability, we constrained the region where the X-rays originate, which is inside the BLR for J0324, but it could not be fully constrained for J2007, although an origin close to the molecular torus is possible. \par
We also tried to estimate the BH mass of these objects by means of the fundamental plane of BH activity. 
However, the blazar nature of our sources prevented us from obtaining any reliable estimate of this parameter. 
We indeed found that $\gamma$-ray-emitting NLS1s and other blazars may follow a different fundamental plane with respect to nonjetted AGN. \par
In conclusion, our analysis showed that $\gamma$-ray NLS1s can be considered as small-scale versions of FSRQs with which they share several physical properties. 
However, further observations of the X-ray properties of jetted and nonjetted NLS1s are needed. 
Broadband data in X-rays by means of current or future facilities like \textit{NuSTAR} and \textit{Athena}, along with multiwavelength observations, are essential to better characterize the whole population.
As a class young AGN, NLS1s are pivotal in shedding light on how AGN and jets are born, and in improving our knowledge of the early phases of feedback between the AGN and its host galaxy.

\begin{acknowledgements}
EP acknowledges financial support under ASI/INAF contract I/037/12/0. This research has made use of the NASA/IPAC Extragalactic Database (NED) which is operated by the Jet Propulsion Laboratory, California Institute of Technology, under contract with the National Aeronautics and  Space Admistration. This research has made use of the \textit{NuSTAR} Data Analysis Software (NuSTARDAS) jointly developed by the ASI Science Data Center (ASDC, Italy) and the California Institute of Technology (USA). This research has made use of the \textit{XRT} Data Analysis Software (XRTDAS) developed under the responsibility of the ASI Science Data Center (ASDC), Italy. Part of this work is based on archival data, software or online services provided by the Space Science Data Center - ASI. This research has made use of the NASA/IPAC Extragalactic Database (NED) which is operated by the Jet Propulsion Laboratory, California Institute of Technology, under contract with the National Aeronautics and Space Administration. 
\end{acknowledgements}

\bibliographystyle{aa}
\bibliography{/home/berton/Desktop/Paper/biblio}

\begin{thebibliography}{129}
\expandafter\ifx\csname natexlab\endcsname\relax\def\natexlab#1{#1}\fi

\bibitem[{{Abdo} {et~al.}(2009{\natexlab{a}}){Abdo}, {Ackermann}, {Ajello},
  {Axelsson}, {Baldini}, {Ballet}, {Barbiellini}, {Bastieri}, {Battelino}, \&
  {Baughman}}]{Abdo09a}
{Abdo}, A.~A., {Ackermann}, M., {Ajello}, M., {et~al.} 2009{\natexlab{a}},
  \apj, 699, 976

\bibitem[{{Abdo} {et~al.}(2009{\natexlab{b}}){Abdo}, {Ackermann}, {Ajello},
  {Axelsson}, {Baldini}, {Ballet}, {Barbiellini}, {Bastieri}, {Baughman},
  {Bechtol}, \& et~al.}]{Abdo09b}
{Abdo}, A.~A., {Ackermann}, M., {Ajello}, M., {et~al.} 2009{\natexlab{b}},
  \apj, 707, 727

\bibitem[{{Abdo} {et~al.}(2009{\natexlab{c}}){Abdo}, {Ackermann}, {Ajello},
  {Baldini}, {Ballet}, {Barbiellini}, {Bastieri}, {Bechtol}, {Bellazzini}, \&
  {Berenji}}]{Abdo09c}
{Abdo}, A.~A., {Ackermann}, M., {Ajello}, M., {et~al.} 2009{\natexlab{c}},
  \apjl, 707, L142

\bibitem[{{Aleksi{\'c}} {et~al.}(2011{\natexlab{a}}){Aleksi{\'c}}, {Antonelli},
  {Antoranz}, {Backes}, {Barrio}, {Bastieri}, {Becerra Gonz{\'a}lez},
  {Bednarek}, {Berdyugin}, {Berger}, {Bernardini}, {Biland}, {Blanch}, {Bock},
  {Boller}, {Bonnoli}, {Borla Tridon}, {Braun}, {Bretz}, {Ca{\~n}ellas},
  {Carmona}, {Carosi}, {Colin}, {Colombo}, {Contreras}, {Cortina}, {Cossio},
  {Covino}, {Dazzi}, {de Angelis}, {de Cea Del Pozo}, {de Lotto}, {Delgado
  Mendez}, {Diago Ortega}, {Doert}, {Dom{\'{\i}}nguez}, {Dominis Prester},
  {Dorner}, {Doro}, {Elsaesser}, {Ferenc}, {Fonseca}, {Font}, {Fruck},
  {Garc{\'{\i}}a L{\'o}pez}, {Garczarczyk}, {Garrido}, {Giavitto},
  {Godinovi{\'c}}, {Hadasch}, {H{\"a}fner}, {Herrero}, {Hildebrand}, {Hose},
  {Hrupec}, {Huber}, {Jogler}, {Klepser}, {Kr{\"a}henb{\"u}hl}, {Krause}, {La
  Barbera}, {Lelas}, {Leonardo}, {Lindfors}, {Lombardi}, {L{\'o}pez}, {Lorenz},
  {Majumdar}, {Makariev}, {Maneva}, {Mankuzhiyil}, {Mannheim}, {Maraschi},
  {Mariotti}, {Mart{\'{\i}}nez}, {Mazin}, {Meucci}, {Miranda}, {Mirzoyan},
  {Miyamoto}, {Mold{\'o}n}, {Moralejo}, {Nieto}, {Nilsson}, {Orito}, {Oya},
  {Paoletti}, {Pardo}, {Paredes}, {Partini}, {Pasanen}, {Pauss},
  {Perez-Torres}, {Persic}, {Peruzzo}, {Pilia}, {Pochon}, {Prada}, {Prada
  Moroni}, {Prandini}, {Puljak}, {Reichardt}, {Reinthal}, {Rhode}, {Rib{\'o}},
  {Rico}, {R{\"u}gamer}, {R{\"u}ger}, {Saggion}, {Saito}, {Saito}, {Salvati},
  {Satalecka}, {Scalzotto}, {Scapin}, {Schultz}, {Schweizer}, {Shayduk},
  {Shore}, {Sillanp{\"a}{\"a}}, {Sitarek}, {Sobczynska}, {Spanier}, {Spiro},
  {Stamerra}, {Steinke}, {Storz}, {Strah}, {Suri{\'c}}, {Takalo}, {Tavecchio},
  {Temnikov}, {Terzi{\'c}}, {Tescaro}, {Teshima}, {Thom}, {Tibolla}, {Torres},
  {Treves}, {Vankov}, {Vogler}, {Wagner}, {Weitzel}, {Zabalza}, {Zandanel}, \&
  {Zanin}}]{Aleksic11a}
{Aleksi{\'c}}, J., {Antonelli}, L.~A., {Antoranz}, P., {et~al.}
  2011{\natexlab{a}}, \aap, 530, A4

\bibitem[{{Aleksi{\'c}} {et~al.}(2011{\natexlab{b}}){Aleksi{\'c}}, {Antonelli},
  {Antoranz}, {Backes}, {Barrio}, {Bastieri}, {Becerra Gonz{\'a}lez},
  {Bednarek}, {Berdyugin}, {Berger}, {Bernardini}, {Biland}, {Blanch}, {Bock},
  {Boller}, {Bonnoli}, {Borla Tridon}, {Braun}, {Bretz}, {Ca{\~n}ellas},
  {Carmona}, {Carosi}, {Colin}, {Colombo}, {Contreras}, {Cortina}, {Cossio},
  {Covino}, {Dazzi}, {De Angelis}, {De Cea del Pozo}, {De Lotto}, {Delgado
  Mendez}, {Diago Ortega}, {Doert}, {Dom{\'{\i}}nguez}, {Dominis Prester},
  {Dorner}, {Doro}, {Elsaesser}, {Ferenc}, {Fonseca}, {Font}, {Fruck},
  {Garc{\'{\i}}a L{\'o}pez}, {Garczarczyk}, {Garrido}, {Giavitto},
  {Godinovi{\'c}}, {Hadasch}, {H{\"a}fner}, {Herrero}, {Hildebrand},
  {H{\"o}hne-M{\"o}nch}, {Hose}, {Hrupec}, {Huber}, {Jogler}, {Klepser},
  {Kr{\"a}henb{\"u}hl}, {Krause}, {La Barbera}, {Lelas}, {Leonardo},
  {Lindfors}, {Lombardi}, {L{\'o}pez}, {Lorenz}, {Makariev}, {Maneva},
  {Mankuzhiyil}, {Mannheim}, {Maraschi}, {Mariotti}, {Mart{\'{\i}}nez},
  {Mazin}, {Meucci}, {Miranda}, {Mirzoyan}, {Miyamoto}, {Mold{\'o}n},
  {Moralejo}, {Nieto}, {Nilsson}, {Orito}, {Oya}, {Paneque}, {Paoletti},
  {Pardo}, {Paredes}, {Partini}, {Pasanen}, {Pauss}, {Perez-Torres}, {Persic},
  {Peruzzo}, {Pilia}, {Pochon}, {Prada}, {Prada Moroni}, {Prandini}, {Puljak},
  {Reichardt}, {Reinthal}, {Rhode}, {Rib{\'o}}, {Rico}, {R{\"u}gamer},
  {Saggion}, {Saito}, {Saito}, {Salvati}, {Satalecka}, {Scalzotto}, {Scapin},
  {Schultz}, {Schweizer}, {Shayduk}, {Shore}, {Sillanp{\"a}{\"a}}, {Sitarek},
  {Sobczynska}, {Spanier}, {Spiro}, {Stamerra}, {Steinke}, {Storz}, {Strah},
  {Suri{\'c}}, {Takalo}, {Tavecchio}, {Temnikov}, {Terzi{\'c}}, {Tescaro},
  {Teshima}, {Thom}, {Tibolla}, {Torres}, {Treves}, {Vankov}, {Vogler},
  {Wagner}, {Weitzel}, {Zabalza}, {Zandanel}, {Zanin}, {MAGIC Collaboration},
  {Tanaka}, {Wood}, \& {Buson}}]{Aleksic11b}
{Aleksi{\'c}}, J., {Antonelli}, L.~A., {Antoranz}, P., {et~al.}
  2011{\natexlab{b}}, \apjl, 730, L8

\bibitem[{{Ant{\'o}n} {et~al.}(2008){Ant{\'o}n}, {Browne}, \&
  {March{\~a}}}]{Anton08}
{Ant{\'o}n}, S., {Browne}, I.~W.~A., \& {March{\~a}}, M.~J. 2008, \aap, 490,
  583

\bibitem[{{Barthelmy} {et~al.}(2005){Barthelmy}, {Barbier}, {Cummings},
  {Fenimore}, {Gehrels}, {Hullinger}, {Krimm}, {Markwardt}, {Palmer},
  {Parsons}, {Sato}, {Suzuki}, {Takahashi}, {Tashiro}, \&
  {Tueller}}]{Barthelmy05}
{Barthelmy}, S.~D., {Barbier}, L.~M., {Cummings}, J.~R., {et~al.} 2005, \ssr,
  120, 143

\bibitem[{{Berton} {et~al.}(2016){Berton}, {Caccianiga}, {Foschini},
  {Peterson}, {Mathur}, {Terreran}, {Ciroi}, {Congiu}, {Cracco}, {Frezzato},
  {La Mura}, \& {Rafanelli}}]{Berton16c}
{Berton}, M., {Caccianiga}, A., {Foschini}, L., {et~al.} 2016, \aap, 591, A98

\bibitem[{{Berton} {et~al.}(2019){Berton}, {Congiu}, {Ciroi}, {Komossa},
  {Frezzato}, {Di Mille}, {Ant{\'o}n}, {Antonucci}, {Caccianiga}, {Coppi},
  {J{\"a}rvel{\"a}}, {Kotilainen}, {L{\"a}hteenm{\"a}ki}, {Mathur}, {Chen},
  {Cracco}, {La Mura}, \& {Rafanelli}}]{Berton19a}
{Berton}, M., {Congiu}, E., {Ciroi}, S., {et~al.} 2019, \aj, 157, 48

\bibitem[{{Berton} {et~al.}(2018{\natexlab{a}}){Berton}, {Congiu},
  {J{\"a}rvel{\"a}}, {Antonucci}, {Kharb}, {Lister}, {Tarchi}, {Caccianiga},
  {Chen}, {Foschini}, {L{\"a}hteenm{\"a}ki}, {Richards}, {Ciroi}, {Cracco},
  {Frezzato}, {La Mura}, \& {Rafanelli}}]{Berton18a}
{Berton}, M., {Congiu}, E., {J{\"a}rvel{\"a}}, E., {et~al.} 2018{\natexlab{a}},
  \aap, 614, A87

\bibitem[{{Berton} {et~al.}(2017){Berton}, {Foschini}, {Caccianiga}, {Ciroi},
  {Congiu}, {Cracco}, {Frezzato}, {La Mura}, \& {Rafanelli}}]{Berton17}
{Berton}, M., {Foschini}, L., {Caccianiga}, A., {et~al.} 2017, Frontiers in
  Astronomy and Space Sciences, 4, 8

\bibitem[{{Berton} {et~al.}(2015){Berton}, {Foschini}, {Ciroi}, {Cracco}, {La
  Mura}, {Lister}, {Mathur}, {Peterson}, {Richards}, \&
  {Rafanelli}}]{Berton15a}
{Berton}, M., {Foschini}, L., {Ciroi}, S., {et~al.} 2015, \aap, 578, A28

\bibitem[{{Berton} {et~al.}(2018{\natexlab{b}}){Berton}, {Liao}, {La Mura},
  {J{\"a}rvel{\"a}}, {Congiu}, {Foschini}, {Frezzato}, {Ramakrishnan}, {Fan},
  {L{\"a}hteenm{\"a}ki}, {Pursimo}, {Abate}, {Bai}, {Calcidese}, {Ciroi},
  {Chen}, {Cracco}, {Li}, {Tornikoski}, \& {Rafanelli}}]{Berton18b}
{Berton}, M., {Liao}, N.~H., {La Mura}, G., {et~al.} 2018{\natexlab{b}}, \aap,
  614, A148

\bibitem[{{Bhatta} {et~al.}(2017){Bhatta}, {Mohorian}, \&
  {Bilinsky}}]{Bhatta17}
{Bhatta}, G., {Mohorian}, M., \& {Bilinsky}, I. 2017, ArXiv e-prints
  [\eprint[arXiv]{1710.09910}]

\bibitem[{{Bhattacharyya} {et~al.}(2014){Bhattacharyya}, {Bhatt}, {Bhatt}, \&
  {Singh}}]{Bhattacharyya14}
{Bhattacharyya}, S., {Bhatt}, H., {Bhatt}, N., \& {Singh}, K.~K. 2014, \mnras,
  440, 106

\bibitem[{{Bianchi} {et~al.}(2009){Bianchi}, {Bonilla}, {Guainazzi}, {Matt}, \&
  {Ponti}}]{Bianchi09}
{Bianchi}, S., {Bonilla}, N.~F., {Guainazzi}, M., {Matt}, G., \& {Ponti}, G.
  2009, \aap, 501, 915

\bibitem[{{Bird} {et~al.}(2007){Bird}, {Malizia}, {Bazzano}, {Barlow},
  {Bassani}, {Hill}, {B{\'e}langer}, {Capitanio}, {Clark}, {Dean}, {Fiocchi},
  {G{\"o}tz}, {Lebrun}, {Molina}, {Produit}, {Renaud}, {Sguera}, {Stephen},
  {Terrier}, {Ubertini}, {Walter}, {Winkler}, \& {Zurita}}]{Bird07}
{Bird}, A.~J., {Malizia}, A., {Bazzano}, A., {et~al.} 2007, \apjs, 170, 175

\bibitem[{{Boroson} \& {Green}(1992)}]{Boroson92}
{Boroson}, T.~A. \& {Green}, R.~F. 1992, \apjs, 80, 109

\bibitem[{{Brandt} {et~al.}(1997){Brandt}, {Mathur}, \& {Elvis}}]{Brandt97}
{Brandt}, W.~N., {Mathur}, S., \& {Elvis}, M. 1997, \mnras, 285, L25

\bibitem[{{Burrows} {et~al.}(2005){Burrows}, {Hill}, {Nousek}, {Kennea},
  {Wells}, {Osborne}, {Abbey}, {Beardmore}, {Mukerjee}, {Short}, {Chincarini},
  {Campana}, {Citterio}, {Moretti}, {Pagani}, {Tagliaferri}, {Giommi},
  {Capalbi}, {Tamburelli}, {Angelini}, {Cusumano}, {Br{\"a}uninger}, {Burkert},
  \& {Hartner}}]{Burrows05}
{Burrows}, D.~N., {Hill}, J.~E., {Nousek}, J.~A., {et~al.} 2005, \ssr, 120, 165

\bibitem[{{Caccianiga} {et~al.}(2015){Caccianiga}, {Ant{\'o}n}, {Ballo},
  {Foschini}, {Maccacaro}, {Della Ceca}, {Severgnini}, {March{\~a}}, {Mateos},
  \& {Sani}}]{Caccianiga15}
{Caccianiga}, A., {Ant{\'o}n}, S., {Ballo}, L., {et~al.} 2015, \mnras, 451,
  1795

\bibitem[{{Caccianiga} {et~al.}(2004){Caccianiga}, {Severgnini}, {Braito},
  {Della Ceca}, {Maccacaro}, {Wolter}, {Barcons}, {Carrera}, {Lehmann}, {Page},
  {Saxton}, \& {Webb}}]{Caccianiga04}
{Caccianiga}, A., {Severgnini}, P., {Braito}, V., {et~al.} 2004, \aap, 416, 901

\bibitem[{{Chaudhury} {et~al.}(2018){Chaudhury}, {Chitnis}, {Rao}, {Singh},
  {Bhattacharyya}, {Dewangan}, {Chakraborty}, {Chandra}, {Stewart}, {Mukerjee},
  \& {Dey}}]{Chaudhury18}
{Chaudhury}, K., {Chitnis}, V.~R., {Rao}, A.~R., {et~al.} 2018, \mnras, 478,
  4830

\bibitem[{{Chen} {et~al.}(2018){Chen}, {Berton}, {La Mura}, {Congiu}, {Cracco},
  {Foschini}, {Fan}, {Ciroi}, {Rafanelli}, \& {Bastieri}}]{Chen18}
{Chen}, S., {Berton}, M., {La Mura}, G., {et~al.} 2018, \aap, 615, A167

\bibitem[{{Chiaberge} \& {Marconi}(2011)}]{Chiaberge11}
{Chiaberge}, M. \& {Marconi}, A. 2011, \mnras, 416, 917

\bibitem[{{Collin} {et~al.}(2006){Collin}, {Kawaguchi}, {Peterson}, \&
  {Vestergaard}}]{Collin06}
{Collin}, S., {Kawaguchi}, T., {Peterson}, B.~M., \& {Vestergaard}, M. 2006,
  \aap, 456, 75

\bibitem[{{Coogan} {et~al.}(2016){Coogan}, {Brown}, \& {Chadwick}}]{Coogan16}
{Coogan}, R.~T., {Brown}, A.~M., \& {Chadwick}, P.~M. 2016, \mnras, 458, 354

\bibitem[{{Costamante} {et~al.}(2018){Costamante}, {Bonnoli}, {Tavecchio},
  {Ghisellini}, {Tagliaferri}, \& {Khangulyan}}]{Costamante18}
{Costamante}, L., {Bonnoli}, G., {Tavecchio}, F., {et~al.} 2018, \mnras, 477,
  4257

\bibitem[{{Dadina}(2007)}]{Dadina07}
{Dadina}, M. 2007, \aap, 461, 1209

\bibitem[{{Dadina}(2008)}]{Dadina08}
{Dadina}, M. 2008, \aap, 485, 417

\bibitem[{{D'Ammando} {et~al.}(2018){D'Ammando}, {Acosta-Pulido}, {Capetti},
  {Baldi}, {Orienti}, {Raiteri}, \& {Ramos Almeida}}]{Dammando18}
{D'Ammando}, F., {Acosta-Pulido}, J.~A., {Capetti}, A., {et~al.} 2018, \mnras,
  478, L66

\bibitem[{{D'Ammando} {et~al.}(2016){D'Ammando}, {Orienti}, {Finke}, {Hovatta},
  {Giroletti}, {Max-Moerbeck}, {Pearson}, {Readhead}, {Reeves}, \&
  {Richards}}]{Dammando16}
{D'Ammando}, F., {Orienti}, M., {Finke}, J., {et~al.} 2016, \mnras, 463, 4469

\bibitem[{{D'Ammando} {et~al.}(2015){D'Ammando}, {Orienti}, {Finke}, {Raiteri},
  {Hovatta}, {Larsson}, {Max-Moerbeck}, {Perkins}, {Readhead}, {Richards},
  {Beilicke}, {Benbow}, {Berger}, {Bird}, {Bugaev}, {Cardenzana}, {Cerruti},
  {Chen}, {Ciupik}, {Dickinson}, {Eisch}, {Errando}, {Falcone}, {Finley},
  {Fleischhack}, {Fortin}, {Fortson}, {Furniss}, {Gerard}, {Gillanders},
  {Griffiths}, {Grube}, {Gyuk}, {H{\aa}kansson}, {Holder}, {Humensky}, {Kar},
  {Kertzman}, {Khassen}, {Kieda}, {Krennrich}, {Kumar}, {Lang}, {Maier},
  {McCann}, {Meagher}, {Moriarty}, {Mukherjee}, {Nieto}, {de Bhr{\'o}ithe},
  {Ong}, {Otte}, {Pohl}, {Popkow}, {Prokoph}, {Pueschel}, {Quinn}, {Ragan},
  {Reynolds}, {Richards}, {Roache}, {Rousselle}, {Santander}, {Sembroski},
  {Smith}, {Staszak}, {Telezhinsky}, {Tucci}, {Tyler}, {Varlotta}, {Vassiliev},
  {Wakely}, {Weinstein}, {Welsing}, {Williams}, \& {Zitzer}}]{Dammando15a}
{D'Ammando}, F., {Orienti}, M., {Finke}, J., {et~al.} 2015, \mnras, 446, 2456

\bibitem[{{Decarli} {et~al.}(2008){Decarli}, {Dotti}, {Fontana}, \&
  {Haardt}}]{Decarli08}
{Decarli}, R., {Dotti}, M., {Fontana}, M., \& {Haardt}, F. 2008, \mnras, 386,
  L15

\bibitem[{{Deo} {et~al.}(2006){Deo}, {Crenshaw}, \& {Kraemer}}]{Deo06}
{Deo}, R.~P., {Crenshaw}, D.~M., \& {Kraemer}, S.~B. 2006, \aj, 132, 321

\bibitem[{{Doi} {et~al.}(2006){Doi}, {Nagai}, {Asada}, {Kameno}, {Wajima}, \&
  {Inoue}}]{Doi06}
{Doi}, A., {Nagai}, H., {Asada}, K., {et~al.} 2006, \pasj, 58, 829

\bibitem[{{Done} {et~al.}(2012){Done}, {Davis}, {Jin}, {Blaes}, \&
  {Ward}}]{Done12}
{Done}, C., {Davis}, S.~W., {Jin}, C., {Blaes}, O., \& {Ward}, M. 2012, \mnras,
  420, 1848

\bibitem[{{Donea} \& {Protheroe}(2003)}]{Donea03}
{Donea}, A.-C. \& {Protheroe}, R.~J. 2003, Astroparticle Physics, 18, 377

\bibitem[{{Foschini}(2011)}]{Foschini11}
{Foschini}, L. 2011, in Narrow-Line Seyfert 1 Galaxies and their Place in the
  Universe, Proc. of Science, Vol. NLS1, id. 24, 24

\bibitem[{{Foschini}(2012)}]{Foschini12}
{Foschini}, L. 2012, in Proceedings of Nuclei of Seyfert galaxies and QSOs -
  Central engine \& conditions of star formation, Proc. of Science, Vol.
  Seyfert 2012, id. 10

\bibitem[{{Foschini}(2014)}]{Foschini14}
{Foschini}, L. 2014, International Journal of Modern Physics Conference Series,
  28, 1460188

\bibitem[{{Foschini}(2017)}]{Foschini17}
{Foschini}, L. 2017, Frontiers in Astronomy and Space Sciences, 4, 6

\bibitem[{{Foschini} {et~al.}(2015){Foschini}, {Berton}, {Caccianiga}, {Ciroi},
  {Cracco}, {Peterson}, {Angelakis}, {Braito}, {Fuhrmann}, {Gallo}, {Grupe},
  {J{\"a}rvel{\"a}}, {Kaufmann}, {Komossa}, {Kovalev}, {L{\"a}hteenm{\"a}ki},
  {Lisakov}, {Lister}, {Mathur}, {Richards}, {Romano}, {Sievers},
  {Tagliaferri}, {Tammi}, {Tibolla}, {Tornikoski}, {Vercellone}, {La Mura},
  {Maraschi}, \& {Rafanelli}}]{Foschini15}
{Foschini}, L., {Berton}, M., {Caccianiga}, A., {et~al.} 2015, \aap, 575, A13

\bibitem[{{Foschini} {et~al.}(2011{\natexlab{a}}){Foschini}, {Ghisellini},
  {Kovalev}, {Lister}, {D'Ammando}, {Thompson}, {Tramacere}, {Angelakis},
  {Donato}, {Falcone}, {Fuhrmann}, {Hauser}, {Kovalev}, {Mannheim}, {Maraschi},
  {Max-Moerbeck}, {Nestoras}, {Pavlidou}, {Pearson}, {Pushkarev}, {Readhead},
  {Richards}, {Stevenson}, {Tagliaferri}, {Tibolla}, {Tavecchio}, \&
  {Wagner}}]{Foschini11c}
{Foschini}, L., {Ghisellini}, G., {Kovalev}, Y.~Y., {et~al.}
  2011{\natexlab{a}}, \mnras, 413, 1671

\bibitem[{{Foschini} {et~al.}(2006){Foschini}, {Ghisellini}, {Raiteri},
  {Tavecchio}, {Villata}, {Maraschi}, {Pian}, {Tagliaferri}, {Di Cocco}, \&
  {Malaguti}}]{Foschini06}
{Foschini}, L., {Ghisellini}, G., {Raiteri}, C.~M., {et~al.} 2006, \aap, 453,
  829

\bibitem[{{Foschini} {et~al.}(2011{\natexlab{b}}){Foschini}, {Ghisellini},
  {Tavecchio}, {Bonnoli}, \& {Stamerra}}]{Foschini11a}
{Foschini}, L., {Ghisellini}, G., {Tavecchio}, F., {Bonnoli}, G., \&
  {Stamerra}, A. 2011{\natexlab{b}}, ArXiv e-prints [\eprint[arXiv]{1110.4471}]

\bibitem[{{Foschini} {et~al.}(2009){Foschini}, {Maraschi}, {Tavecchio},
  {Ghisellini}, {Gliozzi}, \& {Sambruna}}]{Foschini09}
{Foschini}, L., {Maraschi}, L., {Tavecchio}, F., {et~al.} 2009, Advances in
  Space Research, 43, 889

\bibitem[{{Fossati} {et~al.}(1998){Fossati}, {Maraschi}, {Celotti}, {Comastri},
  \& {Ghisellini}}]{Fossati98}
{Fossati}, G., {Maraschi}, L., {Celotti}, A., {Comastri}, A., \& {Ghisellini},
  G. 1998, \mnras, 299, 433

\bibitem[{{Gallo}(2018)}]{Gallo18}
{Gallo}, L. 2018, in Proceedings of Science, vol. Revisiting narrow-line
  Seyfert 1 galaxies and their place in the Universe, 34

\bibitem[{{Gallo}(2006)}]{Gallo06}
{Gallo}, L.~C. 2006, \mnras, 368, 479

\bibitem[{{Gallo} {et~al.}(2006){Gallo}, {Edwards}, {Ferrero}, {Kataoka},
  {Lewis}, {Ellingsen}, {Misanovic}, {Welsh}, {Whiting}, {Boller}, {Brinkmann},
  {Greenhill}, \& {Oshlack}}]{Gallo06a}
{Gallo}, L.~C., {Edwards}, P.~G., {Ferrero}, E., {et~al.} 2006, \mnras, 370,
  245

\bibitem[{{Ghisellini}(2016)}]{Ghisellini16}
{Ghisellini}, G. 2016, Galaxies, 4, 36

\bibitem[{{Ghisellini} {et~al.}(1998){Ghisellini}, {Celotti}, {Fossati},
  {Maraschi}, \& {Comastri}}]{Ghisellini98}
{Ghisellini}, G., {Celotti}, A., {Fossati}, G., {Maraschi}, L., \& {Comastri},
  A. 1998, \mnras, 301, 451

\bibitem[{{Ghisellini} {et~al.}(2010){Ghisellini}, {Tavecchio}, {Foschini},
  {Ghirlanda}, {Maraschi}, \& {Celotti}}]{Ghisellini10}
{Ghisellini}, G., {Tavecchio}, F., {Foschini}, L., {et~al.} 2010, \mnras, 402,
  497

\bibitem[{{Giommi} {et~al.}(1990){Giommi}, {Barr}, {Garilli}, {Maccagni}, \&
  {Pollock}}]{Giommi90}
{Giommi}, P., {Barr}, P., {Garilli}, B., {Maccagni}, D., \& {Pollock}, A.~M.~T.
  1990, \apj, 356, 432

\bibitem[{{Giroletti} \& {Panessa}(2009)}]{Giroletti09}
{Giroletti}, M. \& {Panessa}, F. 2009, \apjl, 706, L260

\bibitem[{{Goodrich}(1989)}]{Goodrich89}
{Goodrich}, R.~W. 1989, \apj, 342, 224

\bibitem[{{Grandi} \& {Palumbo}(2004)}]{Grandi04}
{Grandi}, P. \& {Palumbo}, G. G.~C. 2004, Science, 306, 998

\bibitem[{{Gravity Collaboration} {et~al.}(2018){Gravity Collaboration},
  {Sturm}, {Dexter}, {Pfuhl}, {Stock}, {Davies}, {Lutz}, {Cl{\'e}net},
  {Eckart}, {Eisenhauer}, {Genzel}, {Gratadour}, {H{\"o}nig}, {Kishimoto},
  {Lacour}, {Millour}, {Netzer}, {Perrin}, {Peterson}, {Petrucci}, {Rouan},
  {Waisberg}, {Woillez}, {Amorim}, {Brandner}, {F{\"o}rster Schreiber},
  {Garcia}, {Gillessen}, {Ott}, {Paumard}, {Perraut}, {Scheithauer},
  {Straubmeier}, {Tacconi}, \& {Widmann}}]{Gravity18}
{Gravity Collaboration}, {Sturm}, E., {Dexter}, J., {et~al.} 2018, \nat, 563,
  657

\bibitem[{{Greene} {et~al.}(2010){Greene}, {Hood}, {Barth}, {Bennert}, {Bentz},
  {Filippenko}, {Gates}, {Malkan}, {Treu}, {Walsh}, \& {Woo}}]{Greene10}
{Greene}, J.~E., {Hood}, C.~E., {Barth}, A.~J., {et~al.} 2010, \apj, 723, 409

\bibitem[{{Grupe}(1996)}]{Grupe96}
{Grupe}, D. 1996, PhD thesis, PhD thesis.~Univ.~G{\"o}ttingen , (1996)

\bibitem[{{Grupe}(2000)}]{Grupe00}
{Grupe}, D. 2000, New Astron. Rev., 44, 455

\bibitem[{{Grupe} {et~al.}(2010){Grupe}, {Komossa}, {Leighly}, \&
  {Page}}]{Grupe10}
{Grupe}, D., {Komossa}, S., {Leighly}, K.~M., \& {Page}, K.~L. 2010, \apjs,
  187, 64

\bibitem[{{Grupe} {et~al.}(2004){Grupe}, {Mathur}, \& {Komossa}}]{Grupe04}
{Grupe}, D., {Mathur}, S., \& {Komossa}, S. 2004, \aj, 127, 3161

\bibitem[{{G{\"u}ltekin} {et~al.}(2019){G{\"u}ltekin}, {King}, {Cackett},
  {Nyland}, {Miller}, {Di Matteo}, {Markoff}, \& {Rupen}}]{Gultekin19}
{G{\"u}ltekin}, K., {King}, A.~L., {Cackett}, E.~M., {et~al.} 2019, \apj, 871,
  80

\bibitem[{{Harrison} {et~al.}(2013){Harrison}, {Craig}, {Christensen},
  {Hailey}, {Zhang}, {Boggs}, {Stern}, {Cook}, {Forster}, {Giommi},
  {Grefenstette}, {Kim}, {Kitaguchi}, {Koglin}, {Madsen}, {Mao}, {Miyasaka},
  {Mori}, {Perri}, {Pivovaroff}, {Puccetti}, {Rana}, {Westergaard}, {Willis},
  {Zoglauer}, {An}, {Bachetti}, {Barri{\`e}re}, {Bellm}, {Bhalerao},
  {Brejnholt}, {Fuerst}, {Liebe}, {Markwardt}, {Nynka}, {Vogel}, {Walton},
  {Wik}, {Alexander}, {Cominsky}, {Hornschemeier}, {Hornstrup}, {Kaspi},
  {Madejski}, {Matt}, {Molendi}, {Smith}, {Tomsick}, {Ajello}, {Ballantyne},
  {Balokovi{\'c}}, {Barret}, {Bauer}, {Blandford}, {Brandt}, {Brenneman},
  {Chiang}, {Chakrabarty}, {Chenevez}, {Comastri}, {Dufour}, {Elvis}, {Fabian},
  {Farrah}, {Fryer}, {Gotthelf}, {Grindlay}, {Helfand}, {Krivonos}, {Meier},
  {Miller}, {Natalucci}, {Ogle}, {Ofek}, {Ptak}, {Reynolds}, {Rigby},
  {Tagliaferri}, {Thorsett}, {Treister}, \& {Urry}}]{Harrison13}
{Harrison}, F.~A., {Craig}, W.~W., {Christensen}, F.~E., {et~al.} 2013, \apj,
  770, 103

\bibitem[{{Heinz} \& {Sunyaev}(2003)}]{Heinz03}
{Heinz}, S. \& {Sunyaev}, R.~A. 2003, \mnras, 343, L59

\bibitem[{{Inayoshi} \& {Haiman}(2016)}]{Inayoshi16}
{Inayoshi}, K. \& {Haiman}, Z. 2016, \apj, 828, 110

\bibitem[{{J{\"a}rvel{\"a}} {et~al.}(2018){J{\"a}rvel{\"a}},
  {L{\"a}hteenm{\"a}ki}, \& {Berton}}]{Jarvela18}
{J{\"a}rvel{\"a}}, E., {L{\"a}hteenm{\"a}ki}, A., \& {Berton}, M. 2018, \aap,
  619, A69

\bibitem[{{Kalberla} {et~al.}(2005){Kalberla}, {Burton}, {Hartmann}, {Arnal},
  {Bajaja}, {Morras}, \& {P{\"o}ppel}}]{Kalberla05}
{Kalberla}, P.~M.~W., {Burton}, W.~B., {Hartmann}, D., {et~al.} 2005, \aap,
  440, 775

\bibitem[{{Kalita} {et~al.}(2017){Kalita}, {Gupta}, {Wiita}, {Dewangan}, \&
  {Duorah}}]{Kalita17}
{Kalita}, N., {Gupta}, A.~C., {Wiita}, P.~J., {Dewangan}, G.~C., \& {Duorah},
  K. 2017, \mnras, 469, 3824

\bibitem[{{Komatsu} {et~al.}(2011){Komatsu}, {Smith}, {Dunkley}, {Bennett},
  {Gold}, {Hinshaw}, {Jarosik}, {Larson}, {Nolta}, {Page}, {Spergel},
  {Halpern}, {Hill}, {Kogut}, {Limon}, {Meyer}, {Odegard}, {Tucker}, {Weiland},
  {Wollack}, \& {Wright}}]{Komatsu11}
{Komatsu}, E., {Smith}, K.~M., {Dunkley}, J., {et~al.} 2011, \apjs, 192, 18

\bibitem[{{Komossa}(2018)}]{Komossa18}
{Komossa}, S. 2018, in Proceedings of Science, vol. Revisiting narrow-line
  Seyfert 1 galaxies and their place in the Universe, 15

\bibitem[{{Kotilainen} {et~al.}(2016){Kotilainen}, {Le{\'o}n-Tavares},
  {Olgu{\'{\i}}n-Iglesias}, {Baes}, {An{\'o}rve}, {Chavushyan}, \&
  {Carrasco}}]{Kotilainen16}
{Kotilainen}, J.~K., {Le{\'o}n-Tavares}, J., {Olgu{\'{\i}}n-Iglesias}, A.,
  {et~al.} 2016, \apj, 832, 157

\bibitem[{{Kreikenbohm} {et~al.}(2016){Kreikenbohm}, {Schulz}, {Kadler},
  {Wilms}, {Markowitz}, {Chang}, {Carpenter}, {Els{\"a}sser}, {Gehrels},
  {Mannheim}, {M{\"u}ller}, {Ojha}, {Ros}, \& {Tr{\"u}stedt}}]{Kreikenbohm16}
{Kreikenbohm}, A., {Schulz}, R., {Kadler}, M., {et~al.} 2016, \aap, 585, A91

\bibitem[{{Kynoch} {et~al.}(2018){Kynoch}, {Landt}, {Ward}, {Done}, {Gardner},
  {Boisson}, {Arrieta-Lobo}, {Zech}, {Steenbrugge}, \& {Pereira
  Santaella}}]{Kynoch18}
{Kynoch}, D., {Landt}, H., {Ward}, M.~J., {et~al.} 2018, \mnras, 475, 404

\bibitem[{{L{\"a}hteenm{\"a}ki} {et~al.}(2018){L{\"a}hteenm{\"a}ki},
  {J{\"a}rvel{\"a}}, {Ramakrishnan}, {Tornikoski}, {Tammi}, {Vera}, \&
  {Chamani}}]{Lahteenmaki18}
{L{\"a}hteenm{\"a}ki}, A., {J{\"a}rvel{\"a}}, E., {Ramakrishnan}, V., {et~al.}
  2018, \aap, 614, L1

\bibitem[{{Landt} {et~al.}(2008){Landt}, {Padovani}, {Giommi}, {Perri}, \&
  {Cheung}}]{Landt08}
{Landt}, H., {Padovani}, P., {Giommi}, P., {Perri}, M., \& {Cheung}, C.~C.
  2008, \apj, 676, 87

\bibitem[{{Landt} {et~al.}(2017){Landt}, {Ward}, {Balokovi{\'c}}, {Kynoch},
  {Storchi-Bergmann}, {Boisson}, {Done}, {Schimoia}, \& {Stern}}]{Landt17}
{Landt}, H., {Ward}, M.~J., {Balokovi{\'c}}, M., {et~al.} 2017, \mnras, 464,
  2565

\bibitem[{{Laor}(2000)}]{Laor00}
{Laor}, A. 2000, \apjl, 543, L111

\bibitem[{{Larsson} {et~al.}(2018){Larsson}, {D'Ammando}, {Falocco},
  {Giroletti}, {Orienti}, {Piconcelli}, \& {Righini}}]{Larsson18}
{Larsson}, J., {D'Ammando}, F., {Falocco}, S., {et~al.} 2018, \mnras, 476, 43

\bibitem[{{Laurent-Muehleisen} {et~al.}(1997){Laurent-Muehleisen}, {Kollgaard},
  {Ryan}, {Feigelson}, {Brinkmann}, \& {Siebert}}]{Laurent97}
{Laurent-Muehleisen}, S.~A., {Kollgaard}, R.~I., {Ryan}, P.~J., {et~al.} 1997,
  \aaps, 122 [\eprint{astro-ph/9607058}]

\bibitem[{{Leighly}(1999{\natexlab{a}})}]{Leighly99a}
{Leighly}, K.~M. 1999{\natexlab{a}}, The Astrophysical Journal Supplement
  Series, 125, 297

\bibitem[{{Leighly}(1999{\natexlab{b}})}]{Leighly99b}
{Leighly}, K.~M. 1999{\natexlab{b}}, The Astrophysical Journal Supplement
  Series, 125, 317

\bibitem[{{Le{\'o}n Tavares} {et~al.}(2014){Le{\'o}n Tavares}, {Kotilainen},
  {Chavushyan}, {A{\~n}orve}, {Puerari}, {Cruz-Gonz{\'a}lez},
  {Pati{\~n}o-Alvarez}, {Ant{\'o}n}, {Carrami{\~n}ana}, {Carrasco}, {Guichard},
  {Karhunen}, {Olgu{\'{\i}}n-Iglesias}, {Sanghvi}, \& {Valdes}}]{Leontavares14}
{Le{\'o}n Tavares}, J., {Kotilainen}, J., {Chavushyan}, V., {et~al.} 2014,
  \apj, 795, 58

\bibitem[{{Liu} {et~al.}(2010){Liu}, {Wang}, {Mao}, \& {Wei}}]{Liu10}
{Liu}, H., {Wang}, J., {Mao}, Y., \& {Wei}, J. 2010, \apjl, 715, L113

\bibitem[{{Lubi{\'n}ski} {et~al.}(2016){Lubi{\'n}ski}, {Beckmann}, {Gibaud},
  {Paltani}, {Papadakis}, {Ricci}, {Soldi}, {T{\"u}rler}, {Walter}, \&
  {Zdziarski}}]{Lubinski16}
{Lubi{\'n}ski}, P., {Beckmann}, V., {Gibaud}, L., {et~al.} 2016, \mnras, 458,
  2454

\bibitem[{{Malizia} {et~al.}(2007){Malizia}, {Landi}, {Bassani}, {Bird},
  {Molina}, {De Rosa}, {Fiocchi}, {Gehrels}, {Kennea}, \& {Perri}}]{Malizia07}
{Malizia}, A., {Landi}, R., {Bassani}, L., {et~al.} 2007, \apj, 668, 81

\bibitem[{{Marziani} {et~al.}(2018){Marziani}, {del Olmo}, {D'Onofrio},
  {Dultzin}, {Negrete}, {Mart{\'{\i}}nez-Aldama}, {Bon}, {Bon}, \&
  {Stirpe}}]{Marziani18a}
{Marziani}, P., {del Olmo}, A., {D'Onofrio}, M., {et~al.} 2018, in Proceedings
  of Science, vol. Revisiting narrow-line Seyfert 1 galaxies and their place in
  the Universe, 2

\bibitem[{{Mathur}(2000)}]{Mathur00}
{Mathur}, S. 2000, \mnras, 314, L17

\bibitem[{{Mathur} {et~al.}(2012){Mathur}, {Fields}, {Peterson}, \&
  {Grupe}}]{Mathur12}
{Mathur}, S., {Fields}, D., {Peterson}, B.~M., \& {Grupe}, D. 2012, \apj, 754,
  146

\bibitem[{{Mathur} {et~al.}(2001){Mathur}, {Kuraszkiewicz}, \&
  {Czerny}}]{Mathur01}
{Mathur}, S., {Kuraszkiewicz}, J., \& {Czerny}, B. 2001, \na, 6, 321

\bibitem[{{Merloni} {et~al.}(2003){Merloni}, {Heinz}, \& {di
  Matteo}}]{Merloni03}
{Merloni}, A., {Heinz}, S., \& {di Matteo}, T. 2003, \mnras, 345, 1057

\bibitem[{{Ojha} {et~al.}(2019){Ojha}, {Krishna}, \& {Chand}}]{Ojha19}
{Ojha}, V., {Krishna}, G., \& {Chand}, H. 2019, \mnras, 483, 3036

\bibitem[{{Olgu{\'{\i}}n-Iglesias} {et~al.}(2017){Olgu{\'{\i}}n-Iglesias},
  {Kotilainen}, {Le{\'o}n Tavares}, {Chavushyan}, \&
  {A{\~n}orve}}]{OlguinIglesias17}
{Olgu{\'{\i}}n-Iglesias}, A., {Kotilainen}, J.~K., {Le{\'o}n Tavares}, J.,
  {Chavushyan}, V., \& {A{\~n}orve}, C. 2017, \mnras, 467, 3712

\bibitem[{{Orban de Xivry} {et~al.}(2011){Orban de Xivry}, {Davies},
  {Schartmann}, {Komossa}, {Marconi}, {Hicks}, {Engel}, \&
  {Tacconi}}]{OrbandeXivry11}
{Orban de Xivry}, G., {Davies}, R., {Schartmann}, M., {et~al.} 2011, \mnras,
  417, 2721

\bibitem[{{Orienti} {et~al.}(2012){Orienti}, {D'Ammando}, {Giroletti}, \& {for
  the Fermi-LAT Collaboration}}]{Orienti12}
{Orienti}, M., {D'Ammando}, F., {Giroletti}, M., \& {for the Fermi-LAT
  Collaboration}. 2012, arXiv e-prints [\eprint[arXiv]{1205.0402}]

\bibitem[{{Oshlack} {et~al.}(2001){Oshlack}, {Webster}, \&
  {Whiting}}]{Oshlack01}
{Oshlack}, A.~Y.~K.~N., {Webster}, R.~L., \& {Whiting}, M.~T. 2001, \apj, 558,
  578

\bibitem[{{Osterbrock} \& {Pogge}(1985)}]{Osterbrock85}
{Osterbrock}, D.~E. \& {Pogge}, R.~W. 1985, \apj, 297, 166

\bibitem[{{Padovani} {et~al.}(2002){Padovani}, {Costamante}, {Ghisellini},
  {Giommi}, \& {Perlman}}]{Padovani02}
{Padovani}, P., {Costamante}, L., {Ghisellini}, G., {Giommi}, P., \& {Perlman},
  E. 2002, \apj, 581, 895

\bibitem[{{Paliya} {et~al.}(2019){Paliya}, {Parker}, {Jiang}, {Fabian},
  {Brenneman}, {Ajello}, \& {Hartmann}}]{Paliya19}
{Paliya}, V.~S., {Parker}, M.~L., {Jiang}, J., {et~al.} 2019, \apj, 872, 169

\bibitem[{{Paliya} {et~al.}(2016){Paliya}, {Rajput}, {Stalin}, \&
  {Pandey}}]{Paliya16}
{Paliya}, V.~S., {Rajput}, B., {Stalin}, C.~S., \& {Pandey}, S.~B. 2016, \apj,
  819, 121

\bibitem[{{Paliya} \& {Stalin}(2016)}]{Paliya16a}
{Paliya}, V.~S. \& {Stalin}, C.~S. 2016, \apj, 820, 52

\bibitem[{{Panessa} {et~al.}(2011){Panessa}, {de Rosa}, {Bassani}, {Bazzano},
  {Bird}, {Landi}, {Malizia}, {Miniutti}, {Molina}, \& {Ubertini}}]{Panessa11}
{Panessa}, F., {de Rosa}, A., {Bassani}, L., {et~al.} 2011, \mnras, 417, 2426

\bibitem[{{Peterson}(2011)}]{Peterson11}
{Peterson}, B.~M. 2011, in Narrow-Line Seyfert 1 Galaxies and their Place in
  the Universe, 32

\bibitem[{{Peterson} {et~al.}(2004){Peterson}, {Ferrarese}, {Gilbert}, {Kaspi},
  {Malkan}, {Maoz}, {Merritt}, {Netzer}, {Onken}, {Pogge}, {Vestergaard}, \&
  {Wandel}}]{Peterson04}
{Peterson}, B.~M., {Ferrarese}, L., {Gilbert}, K.~M., {et~al.} 2004, \apj, 613,
  682

\bibitem[{{Piconcelli} {et~al.}(2005){Piconcelli}, {Jimenez-Bail{\'o}n},
  {Guainazzi}, {Schartel}, {Rodr{\'{\i}}guez-Pascual}, \&
  {Santos-Lle{\'o}}}]{Piconcelli05}
{Piconcelli}, E., {Jimenez-Bail{\'o}n}, E., {Guainazzi}, M., {et~al.} 2005,
  \aap, 432, 15

\bibitem[{{Pounds} {et~al.}(1995){Pounds}, {Done}, \& {Osborne}}]{Pounds95}
{Pounds}, K.~A., {Done}, C., \& {Osborne}, J.~P. 1995, \mnras, 277, L5

\bibitem[{{Pushkarev} {et~al.}(2017){Pushkarev}, {Kovalev}, {Lister}, \&
  {Savolainen}}]{Pushkarev17}
{Pushkarev}, A.~B., {Kovalev}, Y.~Y., {Lister}, M.~L., \& {Savolainen}, T.
  2017, \mnras, 468, 4992

\bibitem[{{Rakshit} {et~al.}(2017){Rakshit}, {Stalin}, {Chand}, \&
  {Zhang}}]{Rakshit17a}
{Rakshit}, S., {Stalin}, C.~S., {Chand}, H., \& {Zhang}, X.-G. 2017, \apjs,
  229, 39

\bibitem[{{Reeves} \& {Turner}(2000)}]{Reeves00}
{Reeves}, J.~N. \& {Turner}, M.~J.~L. 2000, \mnras, 316, 234

\bibitem[{{Romano} {et~al.}(2018){Romano}, {Vercellone}, {Foschini},
  {Tavecchio}, {Landoni}, \& {Kn{\"o}dlseder}}]{Romano18}
{Romano}, P., {Vercellone}, S., {Foschini}, L., {et~al.} 2018, \mnras, 481,
  5046

\bibitem[{{Schulz} {et~al.}(2016){Schulz}, {Kreikenbohm}, {Kadler}, {Ojha},
  {Ros}, {Stevens}, {Edwards}, {Carpenter}, {Els{\"a}sser}, {Gehrels},
  {Gro{\ss}berger}, {Hase}, {Horiuchi}, {Lovell}, {Mannheim}, {Markowitz},
  {M{\"u}ller}, {Phillips}, {Pl{\"o}tz}, {Quick}, {Tr{\"u}stedt}, {Tzioumis},
  \& {Wilms}}]{Schulz15}
{Schulz}, R., {Kreikenbohm}, A., {Kadler}, M., {et~al.} 2016, \aap, 588, A146

\bibitem[{{Shen} \& {Ho}(2014)}]{Shen14}
{Shen}, Y. \& {Ho}, L.~C. 2014, \nat, 513, 210

\bibitem[{{Str{\"u}der} {et~al.}(2001){Str{\"u}der}, {Briel}, {Dennerl},
  {Hartmann}, {Kendziorra}, {Meidinger}, {Pfeffermann}, {Reppin}, {Aschenbach},
  {Bornemann}, {Br{\"a}uninger}, {Burkert}, {Elender}, {Freyberg}, {Haberl},
  {Hartner}, {Heuschmann}, {Hippmann}, {Kastelic}, {Kemmer}, {Kettenring},
  {Kink}, {Krause}, {M{\"u}ller}, {Oppitz}, {Pietsch}, {Popp}, {Predehl},
  {Read}, {Stephan}, {St{\"o}tter}, {Tr{\"u}mper}, {Holl}, {Kemmer}, {Soltau},
  {St{\"o}tter}, {Weber}, {Weichert}, {von Zanthier}, {Carathanassis}, {Lutz},
  {Richter}, {Solc}, {B{\"o}ttcher}, {Kuster}, {Staubert}, {Abbey}, {Holland},
  {Turner}, {Balasini}, {Bignami}, {La Palombara}, {Villa}, {Buttler},
  {Gianini}, {Lain{\'e}}, {Lumb}, \& {Dhez}}]{Struder01}
{Str{\"u}der}, L., {Briel}, U., {Dennerl}, K., {et~al.} 2001, \aap, 365, L18

\bibitem[{{Sulentic} {et~al.}(2000){Sulentic}, {Zwitter}, {Marziani}, \&
  {Dultzin-Hacyan}}]{Sulentic00}
{Sulentic}, J.~W., {Zwitter}, T., {Marziani}, P., \& {Dultzin-Hacyan}, D. 2000,
  \apjl, 536, L5

\bibitem[{{Tavecchio} {et~al.}(2013){Tavecchio}, {Pacciani}, {Donnarumma},
  {Stamerra}, {Isler}, {MacPherson}, \& {Urry}}]{Tavecchio13}
{Tavecchio}, F., {Pacciani}, L., {Donnarumma}, I., {et~al.} 2013, \mnras, 435,
  L24

\bibitem[{{Titarchuk}(1994)}]{Titarchuk94}
{Titarchuk}, L. 1994, \apj, 434, 570

\bibitem[{{Tortosa} {et~al.}(2018){Tortosa}, {Bianchi}, {Marinucci}, {Matt}, \&
  {Petrucci}}]{Tortosa18}
{Tortosa}, A., {Bianchi}, S., {Marinucci}, A., {Matt}, G., \& {Petrucci}, P.~O.
  2018, \aap, 614, A37

\bibitem[{{Turner} {et~al.}(2001){Turner}, {Abbey}, {Arnaud}, {Balasini},
  {Barbera}, {Belsole}, {Bennie}, {Bernard}, {Bignami}, {Boer}, {Briel},
  {Butler}, {Cara}, {Chabaud}, {Cole}, {Collura}, {Conte}, {Cros}, {Denby},
  {Dhez}, {Di Coco}, {Dowson}, {Ferrando}, {Ghizzardi}, {Gianotti}, {Goodall},
  {Gretton}, {Griffiths}, {Hainaut}, {Hochedez}, {Holland}, {Jourdain},
  {Kendziorra}, {Lagostina}, {Laine}, {La Palombara}, {Lortholary}, {Lumb},
  {Marty}, {Molendi}, {Pigot}, {Poindron}, {Pounds}, {Reeves}, {Reppin},
  {Rothenflug}, {Salvetat}, {Sauvageot}, {Schmitt}, {Sembay}, {Short},
  {Spragg}, {Stephen}, {Str{\"u}der}, {Tiengo}, {Trifoglio}, {Tr{\"u}mper},
  {Vercellone}, {Vigroux}, {Villa}, {Ward}, {Whitehead}, \& {Zonca}}]{Turner01}
{Turner}, M.~J.~L., {Abbey}, A., {Arnaud}, M., {et~al.} 2001, \aap, 365, L27

\bibitem[{{Turner} {et~al.}(2018){Turner}, {Reeves}, {Braito}, \&
  {Costa}}]{Turner18}
{Turner}, T.~J., {Reeves}, J.~N., {Braito}, V., \& {Costa}, M. 2018, \mnras,
  476, 1258

\bibitem[{{Valtaoja} {et~al.}(1999){Valtaoja}, {L{\"a}hteenm{\"a}ki},
  {Ter{\"a}sranta}, \& {Lainela}}]{Valtaoja99}
{Valtaoja}, E., {L{\"a}hteenm{\"a}ki}, A., {Ter{\"a}sranta}, H., \& {Lainela},
  M. 1999, \apjs, 120, 95

\bibitem[{{Wang} {et~al.}(2016){Wang}, {Du}, {Hu}, {Bai}, {Wang}, {Yi}, {Wang},
  {Zhang}, {Xin}, {Lun}, {Chang}, \& {Fan}}]{Wang16}
{Wang}, F., {Du}, P., {Hu}, C., {et~al.} 2016, \apj, 824, 149

\bibitem[{{Younes} {et~al.}(2019){Younes}, {Ptak}, {Ho}, {Xie}, {Terasima},
  {Yuan}, {Huppenkothen}, \& {Yukita}}]{Younes19}
{Younes}, G., {Ptak}, A., {Ho}, L.~C., {et~al.} 2019, \apj, 870, 73

\bibitem[{{Yuan} {et~al.}(2008){Yuan}, {Zhou}, {Komossa}, {Dong}, {Wang}, {Lu},
  \& {Bai}}]{Yuan08}
{Yuan}, W., {Zhou}, H.~Y., {Komossa}, S., {et~al.} 2008, \apj, 685, 801

\bibitem[{{Zhang} {et~al.}(2018){Zhang}, {Zhang}, \& {Zhang}}]{Zhang18}
{Zhang}, X., {Zhang}, H., \& {Zhang}, X. 2018, \apss, 363, 259

\bibitem[{{Zhou} {et~al.}(2007){Zhou}, {Wang}, {Yuan}, {Shan}, {Komossa}, {Lu},
  {Liu}, {Xu}, {Bai}, \& {Jiang}}]{Zhou07}
{Zhou}, H., {Wang}, T., {Yuan}, W., {et~al.} 2007, \apjl, 658, L13

\bibitem[{{Zhou} \& {Wang}(2002)}]{Zhou02}
{Zhou}, H.-Y. \& {Wang}, T.-G. 2002, Ch.J. A\&A, 2, 501

\bibitem[{{Zhou} {et~al.}(2003){Zhou}, {Wang}, {Dong}, {Zhou}, \&
  {Li}}]{Zhou03}
{Zhou}, H.-Y., {Wang}, T.-G., {Dong}, X.-B., {Zhou}, Y.-Y., \& {Li}, C. 2003,
  \apj, 584, 147

\end{thebibliography}

\begin{appendix}
\section{Tables}

\begin{table*}[!ht]
\caption{Summary of \textit{NuSTAR} observations.}
\centering
\footnotesize
\begin{tabular}{l c c c c c c}
\hline\hline
Source & Obsid & Date & Exposure & Net flux FPMA  & Net flux FPMB \\
\hline
J0324 & 60061360002 & 2014-03-15 & 101.6 & 1.930$\pm$0.014 & 1.937$\pm$0.014 \\
J0948 & 60201052002 & 2016-11-04 & 192.7 & 0.396$\pm$0.005 & 0.377$\pm$0.005 \\
J1505 & 60201044002 & 2017-02-12 & 115.4 & 0.090$\pm$0.004 & 0.077$\pm$0.004 \\
J2007 & 80201024002 & 2016-05-09 & 49.2 & 0.224$\pm$0.009 & 0.206$\pm$0.009 \\
J2007 & 60201045002 & 2016-10-23 & 60.6 & 0.268$\pm$0.008 & 0.226$\pm$0.008 \\
\hline\hline 
\end{tabular} 
\tablefoot{Columns: (1) Source; (2) obsid; (3) observation date; (4) exposure time (in ks); (5) net flux in FPMA (in units of 0.1 counts per second); (6) net flux in FPMB (in units of 0.1 counts per second).}
\label{tab:obs_nustar}
\end{table*}

\begin{table*}[!ht]

\caption{Summary of XMM-Netwon observations.}
\centering
\footnotesize
\begin{tabular}{l c c c c c}
\hline\hline
Source & Instrument & Obsid & Date & Exposure (ks) & Net flux (cps) \\
\hline
J0948 & pn & 0790860101 & 2016-11-04 & 63.9 & 0.61$\pm$0.03 \\
J0948 & MOS1 & 0790860101 & 2016-11-04 & 84.9 & 0.14$\pm$0.01 \\
J0948 & MOS2 & 0790860101 & 2016-11-04 & 87.9 & 0.16$\pm$0.01 \\
J2007 & pn & 0790630101 & 2016-05-05 & 53.4 & 0.28$\pm$0.03 \\
J2007 & MOS1 & 0790630101 & 2016-05-05 & 39.7 & 0.07$\pm$0.01 \\
J2007 & MOS2 & 0790630101 & 2016-05-05 & 39.6 & 0.07$\pm$0.01 \\
\hline\hline 
\end{tabular} 
\tablefoot{Columns: (1) Source; (2) instrument; (3) obsid; (4) observation date; (5) exposure time (in ks); (6) net flux (in counts per second).}
\label{tab:obs_xmm}
\end{table*}

\begin{table*}

\caption{Summary of \textit{Swift/XRT} observations.}
\centering
\footnotesize
\begin{tabular}{l c c c }
\hline\hline
Obsid	& Date	& Exposure (ks) & Net flux (cps) \\
\hline
\textbf{J0324} \\
00036533038 & 2013-07-12 & 2.7 & \\
00036533039 & 2013-07-16 & 2.8 & \\
00036533040 & 2013-07-18 & 3.7 & \\
00036533041 & 2013-07-19 & 3.9 & \\
00036533042 & 2013-07-19 & 3.9 & \\
00036533043 & 2013-08-19 & 3.7 & \\
00036533044 & 2013-08-20 & 3.9 & \\
00036533045 & 2013-08-21 & 3.9 & \\
00036533046 & 2013-08-30 & 2.0 & \\
00036533047 & 2013-09-06 & 2.5 & \\
00036533048 & 2013-09-13 & 1.9 & \\
00036533049 & 2013-09-20 & 2.6 & \\
00036533050 & 2013-09-27 & 1.6 & \\
00036533051 & 2013-10-02 & 1.0 & \\
00036533052 & 2014-12-10 & 3.0 & \\
00036533053 & 2014-12-12 & 2.9 & \\
\textit{Total} & & 46.0 & 0.19$\pm$0.01 \\
\hline
\textbf{J1505} \\
00031445002 & 2015-12-21 & 2.8 & \\
00031445003 & 2015-12-25 & 2.2 & \\
00031445004 & 2016-01-01 & 2.7 & \\
00031445005 & 2016-01-08 & 2.2 & \\
00031445006 & 2016-01-14 & 2.7 & \\
00031445007 & 2016-01-22 & 0.8 & \\
00031445008 & 2016-06-06 & 0.7 & \\
00031445009 & 2016-06-14 & 1.9 & \\
00031445010 & 2016-06-24 & 2.2 & \\
00081880001 & 2017-02-12 & 1.8 & \\
\textit{Total} &  & 20 & (9.53$\pm$0.70)$\times$10$^{-3}$ \\
\hline\hline 
\end{tabular} 
\tablefoot{Columns: (1) Obsid; (2) observation date; (3) exposure time (in ks); (4) net flux (in counts per second), esimated only for the total exposure.}
\label{tab:obs_swift}
\end{table*}

\begin{sidewaystable*}
\caption{Results of the spectral fitting.}
\label{tab:results}
\centering
\scalebox{0.75}{
\footnotesize
\begin{tabular}{l c c c c c c c c c c c c c c c c c} 
\hline\hline
Model & $\Gamma_1$ & E$_{break}$ & $\Gamma_2$ & norm$_1$ & T$_0$ & kT & $\tau_p$ & Nh & norm$_2$ & c$_1$ & c$_2$ & c$_3$ & c$_4$ & c$_5$ & c$_6$ & $\chi^2_\nu$/dof \\
{} & {} & {} & {} & {} & $\sigma_{\rm line}$ & EW & $\log{\xi}$ & {} & {} & {} & {} & {} & {} & {} & {} \\
\hline
\textbf{J0324} & {} & {} & {} & {} & {} & {} & {} & {} & {} & FPMB & {} & \\
\hline
TBabs*zpo & 1.80$\pm$0.02 & {} & {} & (2.64$\pm$0.10)$\times 10^{-3}$ & {} & {} & {} & {} & {} & 1.05$\pm$0.03 & {} & {} & {} & {} & {} & 1.14/318 \\
TBabs*bknpo & 1.83$\pm$0.02 & 13.40$^{+3.39}_{-3.86}$ & 1.68$\pm$0.07 & (2.50$\pm$0.11)$\times 10^{-3}$ & {} & {} & {} & {} & {} & 1.05$\pm$0.02 & {} & {} & {} & {} & {} & 1.11/316 \\
TBabs*(zpo+compTT) & 1.67$\pm$0.10 & {} & {} & (1.78$\pm$$^{+0.45}_{-0.62}$)$\times 10^{-3}$ & 30$^{+10}_{-12}$ & $\leq$5.27 & 6.24$^{+0.92}_{-1.90}$ & {} & (7.15$^{+1.33}_{-3.92}$)$\times 10^{-3}$ & 1.05$\pm$0.02 & {} & {} & {} & {} & {} & 1.12/315 \\
TBabs*cutoffpl & 1.78$\pm$0.02 & $\leq$349 & {} & (2.32$\pm$0.09)$\times 10^{-3}$ & {} & {} & {} & {} & {} & 1.05$\pm$0.02 & {} & {} & {} & {} & {} & 1.16/317\\
TBabs*(bknpo+zgauss) & 1.83$\pm$0.03 & 13.01$^{+3.16}_{-4.36}$ & 1.68$\pm$0.07 & (2.49$^{+0.16}_{-0.11}$)$\times 10^{-3}$ & $\leq$1.65 & $\leq$0.03 & {} & {} & (2.70$\pm$2.20)$\times 10^{-6}$ & 1.03$\pm$0.02 & {} & {} & {} & {} & {} & 1.11/314 \\
TBabs*(zpo+zbbody) & 1.72$\pm$0.05 & {} & {} &  (2.14$\pm$0.28)$\times 10^{-3}$ & {} & 1.12$^{+0.17}_{-0.24}$ & {} & {} & (9.74$\pm$0.55)$\times 10^{-6}$ & 1.05$\pm$0.02 & {} & {} & {} & {} & {} & 1.12/316 \\
TBabs*bknpo*zxipcf & 1.84$\pm$0.05 & 13.18$^{+3.31}_{-4.02}$ & 1.68$\pm$0.07 & (2.55$\pm$0.18)$\times 10^{-3}$ & {} & {} & $\geq$-1.74 & $\leq$2.39 & {} & 1.05$\pm$0.02 & {} & {} & {} & {} & {} & 1.12/314 \\
\hline
\textbf{J0948} & {} & {} & {} & {} & {} & {} & {} & {} & {} & MOS1 & MOS2 & FPMA & FPMB \\
\hline
TBabs*zpo & 1.71$\pm$0.01 & {} & {} & (7.11$\pm$0.08)$\times$10$^{-4}$ & {} & {} & {} & {} & {} & 1.00$\pm$0.02 & 0.99$\pm$0.02 & 1.22$\pm$0.04 & 1.23$\pm$0.04 & {} & {} & 1.79/1222 \\
TBabs*bknpo & 1.98$\pm$0.03 & 1.79$\pm$0.13 & 1.41$\pm$0.03 & (3.14$\pm$0.05)$\times 10^{-4}$ & {} & {} & {} & {} & {} & 1.02$\pm$0.02 & 1.02$\pm$0.02 & 0.96$\pm$0.03 & 0.98$\pm$0.03 & {} & {} & 1.06/1220 \\
TBabs*(zpo+compTT) & 1.31$\pm$0.04 & {} & {} & (3.19$^{+0.33}_{-0.17}$)$\times 10^{-4}$ & $\leq$81 & $\geq$2.1 & $\leq$5.2 & {} & $\leq$3.6$\times 10^{-2}$ & 1.02$\pm$0.02 & 1.01$\pm$0.02 & 0.96$\pm$0.02 & 0.98$\pm$0.02 & {} & {} & 1.01/1218 \\
TBabs*cutoffpl & 1.71$\pm$0.01 & $\geq$468 & {} & (3.23$\pm$0.03)$\times$10$^{-4}$ & & {} & {} & {} & {} & 1.00$\pm$0.02 & 0.99$\pm$0.02 & 1.23$\pm$0.04 & 1.24$\pm$0.04 & {} & {} & 1.87/1221 \\
TBabs*(bknpo+zgauss) & 1.98$\pm$0.02 & 1.79$\pm$0.12 & 1.42$\pm$0.02 &  (3.14$\pm$0.35)$\times 10^{-4}$ & $\leq$0.2 & $\leq$0.017 & {} & {} & $\leq$8.77$\times 10^{-7}$ & 1.02$\pm$0.02 & 1.01$\pm$0.02 & 0.96$\pm$0.04 & 0.98$\pm$0.04 & {} & {} & 1.06/1218 \\
TBabs*(zpo+zbbody) & 1.50$\pm$0.02 & {} & {} &  (5.14$\pm$0.14)$\times 10^{-4}$ & {} & 0.18$\pm$0.01 & {} & {} & (9.70$\pm$0.51)$\times 10^{-6}$ & 1.02$\pm$0.02 & 1.01$\pm$0.02 & 1.00$\pm$0.03 & 1.02$\pm$0.03 & {} & {} & 1.12/1220 \\
TBabs*bknpo*zxipcf & 1.93$\pm$0.02 & 2.33$\pm$0.18 & 1.42$\pm$0.03 & (3.68$\pm$0.13)$\times 10^{-4}$ & {} & {} & 3.31$\pm$0.06 & 10.3$\pm$4.4 & {} & 1.02$\pm$0.02 & 1.01$\pm$0.02 & 0.95$\pm$0.03 & 0.97$\pm$0.03 & {} & {} & 1.06/1218 \\
\hline
\textbf{J1505} & {} & {} & {} & {} & {} & {} & {} &{} & {} & FPMA & FPMB \\
\hline
TBabs*zpo & 1.16$\pm$0.08 & {} & {} & (8.57$^{+1.38}_{-1.28}$)$\times$10$^{-5}$ & & {} & {} & {} &{} & 0.55$\pm$0.10 & 0.52$\pm$0.10 & {} & {} & {} & {} & 1.26/68 \\
TBabs*bknpo & 1.40$\pm$0.20 & 5.45$^{+2.97}_{-2.07}$ & 1.08$\pm$0.10 & (6.71$\pm$1.05)$\times 10^{-5}$ & {} & {} & {} &{} & {} & 0.68$^{+0.23}_{-0.15}$ & 0.65$^{+0.22}_{-0.15}$ & {} & {} & {} & {} & 1.22/66\\
TBabs*(zpo+compTT) & 1.32$^{+0.16}_{-0.19}$ & {} & {} & (9.48$^{+1.64}_{-1.89}$)$\times$10$^{-5}$ & $\leq$9331 & $\leq$8.03 & $\leq$6.28 & $\geq$9.1$\times10^{-7}$ & {} &0.66$^{+0.19}_{-0.14}$ & 0.62$^{+0.19}_{-0.13}$ & {} & {} & {} & {} & 1.22/64 \\
TBabs*cutoffpl & 1.14$^{+0.09}_{-0.12}$ & $\geq$86 & {} & (5.71$^{+0.82}_{-0.78}$)$\times$10$^{-5}$ & {} & {} & {} & {} &{} & 0.54$^{+0.09}_{-0.11}$ & 0.52$^{+0.09}_{-0.11}$ & {} & {} & {} & {} & 1.28/67 \\
\hline
\textbf{J2007} & {} & {} & {} & {} & {} & {} & {} & {} &{} & MOS1 & MOS2 & FPMA & FPMB & FPMA & FPMB \\
\hline
TBabs*zpo & 1.62$\pm$0.02 & {} & {} & (1.88$\pm$0.04)$\times 10^{-4}$ & {} & {} & {} & {} & {} & 1.06$\pm$0.04 & 1.01$\pm$0.04 & 1.41$\pm$0.10 & 1.42$\pm$0.11 & 1.65$\pm$0.10 & 1.52$\pm$0.10 & 1.01/377 \\
TBabs*bknpo & 1.66$\pm$0.04 & 2.02$^{+1.11}_{-0.60}$ & 1.51$\pm$0.05 & (1.32$\pm0.27)\times 10^{-4}$ & {} & {} & {} & {} & {} & 1.07$\pm$0.04 & 1.01$\pm$0.04 & 1.34$\pm$0.10 & 1.36$\pm$0.11 & 1.58$\pm$0.10 & 1.45$\pm$0.10 & 0.99/375 \\
TBabs*(zpo+compTT) & 1.60$^{+0.03}_{-0.05}$ & {} & {} & (1.83$^{+0.05}_{-0.33}$)$\times 10^{-4}$ & 9.9$^{+39.5}_{-9.0}$ & 21.5$^{+54.2}_{-16.1}$ & $\leq$75.1 & $\leq$112.5 &{} & 1.06$\pm$0.04 & 1.01$\pm$0.04 & 1.38$\pm$0.10 & 1.39$\pm$0.11 & 1.62$\pm$0.10 & 1.49$\pm$0.10 & 0.99/373 \\
TBabs*cutoffpl & 1.61$\pm$0.02 & $\geq$199 & {} & (1.33$\pm$0.25)$\times 10^{-4}$ & {} & {} & {} & {} &{} & 1.06$\pm$0.04 & 1.01$\pm$0.04 & 1.41$\pm$0.10 & 1.43$\pm$0.11 & 1.66$\pm$0.10 & 1.53$\pm$0.10 & 1.01/376\\
TBabs*(bknpo+zgauss) & 1.67$\pm$0.02 & 2.89$^{+3.19}_{-1.01}$ & 1.54$\pm$0.05 & (1.31$\pm$0.26)$\times 10^{-4}$ & 1.57$^{+1.09}_{-0.98}$ & $\leq$0.34 & {} & {} & (4.01$^{+4.03}_{-2.41}$)$\times 10^{-6}$ & 1.07$\pm$0.04 & 1.02$\pm$0.04 & 1.33$\pm$0.10 & 1.34$\pm$0.11 & 1.56$\pm$0.10 & 1.43$\pm$0.10 & 0.97/373 \\
TBabs*(zpo+zbbody) & 1.60$\pm$0.02 & {} & {} &  (1.83$\pm$0.04)$\times 10^{-4}$ & {} & 5.25$^{+3.09}_{-2.47}$ & {} & {} & (4.06$\pm$3.75)$\times 10^{-6}$ & 1.06$\pm$0.04 & 1.01$\pm$0.04 & 1.38$\pm$0.10 & 1.39$\pm$0.11 & 1.62$\pm$0.10 & 1.49$\pm$0.10 & 0.99/375 \\
TBabs*bknpo*zxipcf & 1.95$^{+0.27}_{-0.21}$ & 1.64$^{+0.69}_{-0.15}$ & 1.64$\pm$0.05 & (1.52$^{+0.21}_{-0.13}$)$\times 10^{-4}$ & {} & {} & -0.11$^{+1.35}_{-0.39}$ & $\leq$0.17 & {} & 1.06$\pm$0.04 & 1.01$\pm$0.04 & 1.34$\pm$0.10 & 1.36$\pm$0.11 & 1.58$\pm$0.10 & 1.45$\pm$0.10 & 0.97/373 \\
\hline\hline
\end{tabular}
}
\tablefoot{Columns: (1) model; (2) photon index; (3) break energy (keV) for broken power law and cut-off energy for cut-off power law; (4) second photon index for broken power law; (5) normalization; (6) soft photon temperature or sigma of the Fe K$\alpha$ line (eV); (7) plasma temperature or equivalent width of Fe K$\alpha$ (keV); (8) optical depth of the plasma or logarithm of the ionization parameter (in erg cm s$^{-1}$); (9) column density of warm absorber ($\times 10^{22}$ cm$^{-2}$); (10) normalization of CompTT; (11),(12),(13),(14),(15),(16) cross-calibration constants for each instrument; (17) $\chi^2_\nu$ and degrees of freedom.}
\end{sidewaystable*}

\clearpage
\begin{table*}
\caption{Observed fluxes and intrinsic luminosity for different models.}
\label{tab:fluxes}
\centering
\footnotesize
\begin{tabular}{l c c c c c c c c} 
\hline\hline
Model & Flux$^o_{0.3-10}$ & logL$^i_{0.3-10}$ & Flux$^o_{2-10}$ & logL$^i_{2-10}$ & Flux$^o_{0.3-70}$ & logL$^i_{0.3-70}$ \\
\hline
\textbf{J0324} \\
\hline
TBabs*zpo & 		12.2$\pm$0.5 & 44.09$\pm$0.03 & 8.2$\pm$0.2 & 43.82$\pm$0.02 & 26.6$\pm$0.7 & 44.37$\pm$0.03 \\
TBabs*bknpo & 		12.5$\pm$0.5 & 44.10$\pm$0.04 & 8.3$\pm$0.4 & 43.83$\pm$0.05 & 27.8$\pm$1.5 & 44.38$\pm$0.06 \\
TBabs*(zpo+compTT) & 	12.3$\pm$0.6 & 44.10$\pm$0.05 & 8.2$\pm$0.4 & 43.83$\pm$0.05 & 27.7$\pm$1.3 & 44.40$\pm$0.05 \\
TBabs*cutoffpl & 	12.1$\pm$0.4 & 44.08$\pm$0.03 & 8.2$\pm$0.3 & 43.82$\pm$0.04 & 26.1$\pm$0.8 & 44.37$\pm$0.03 \\
TBabs*(bknpo+zgauss) & 	12.2$\pm$0.5 & 44.08$\pm$0.03 & 8.2$\pm$0.4 & 43.82$\pm$0.05 & 26.5$\pm$1.0 & 44.37$\pm$0.04 \\
TBabs*(zpo+zbbody) & 	11.5$\pm$1.4 & 44.05$\pm$0.12 & 8.1$\pm$1.0 & 43.82$\pm$0.12 & 26.5$\pm$3.2 & 44.37$\pm$0.12 \\
TBabs*bknpo*zxipcf & 	11.6$\pm$1.1 & 44.03$\pm$0.10 & 8.2$\pm$0.8 & 43.82$\pm$0.10 & 26.9$\pm$2.6 & 44.36$\pm$0.10 \\
\hline
\textbf{J0948} \\
\hline

TBabs*zpo & 		1.99$\pm$0.02 & 45.15$\pm$0.02 & 1.28$\pm$0.03 & 44.92$\pm$0.02 & 4.59$\pm$0.10 & 45.49$\pm$0.02 \\
TBabs*bknpo & 		2.15$\pm$0.03 & 45.13$\pm$0.01 & 1.45$\pm$0.02 & 44.91$\pm$0.01 & 7.18$\pm$0.09 & 45.62$\pm$0.01 \\
TBabs*(zpo+compTT) & 	2.15$\pm$0.22 & 45.27$\pm$0.10 & 1.43$\pm$0.14 & 45.05$\pm$0.10 & 7.74$\pm$0.77 & 45.79$\pm$0.10 \\
TBabs*cutoffpl & 	1.99$\pm$0.02 & 45.15$\pm$0.01 & 1.28$\pm$0.01 & 44.91$\pm$0.01 & 4.44$\pm$0.04 & 45.47$\pm$0.01 \\
TBabs*(bknpo+zgauss) & 	2.15$\pm$0.24 & 45.13$\pm$0.11 & 1.45$\pm$0.16 & 44.92$\pm$0.11 & 7.22$\pm$0.80 & 45.62$\pm$0.11 \\
TBabs*(zpo+zbbody) & 	2.14$\pm$0.28 & 45.23$\pm$0.13 & 1.44$\pm$0.19 & 45.01$\pm$0.13 & 6.47$\pm$0.85 & 45.68$\pm$0.13 \\
Tbabs*bknpo*zxipcf & 	2.15$\pm$0.08 & 45.14$\pm$0.04 & 1.44$\pm$0.05 & 44.92$\pm$0.03 & 7.44$\pm$0.26 & 45.65$\pm$0.04 \\

\hline
\textbf{J1505} \\
\hline
TBabs*zpo & 		0.70$\pm$0.11 & 44.45$\pm$0.17 & 0.56$\pm$0.09 & 44.35$\pm$0.16 & 3.86$\pm$0.62 & 45.19$\pm$0.16 \\
TBabs*bknpo & 		0.62$\pm$0.10 & 44.37$\pm$0.15 & 0.47$\pm$0.07 & 44.23$\pm$0.15 & 3.43$\pm$0.54 & 45.10$\pm$0.16 \\
TBabs*(zpo+compTT) & 	0.64$\pm$0.11 & 44.38$\pm$0.17 & 0.52$\pm$0.09 & 44.30$\pm$0.17 & 2.95$\pm$0.51 & 45.04$\pm$0.17 \\
TBabs*cutoffpl & 	0.64$\pm$0.09 & 44.41$\pm$0.14 & 0.57$\pm$0.08 & 44.36$\pm$0.14 & 2.95$\pm$0.40 & 45.07$\pm$0.14 \\
\hline
\textbf{J2007} \\
\hline
TBabs*zpo & 		0.93$\pm$0.01 & 44.08$\pm$0.01 & 0.61$\pm$0.01 & 43.87$\pm$0.02 & 2.40$\pm$0.03 & 44.48$\pm$0.01 \\
TBabs*bknpo & 		0.94$\pm$0.19 & 44.09$\pm$0.20 & 0.62$\pm$0.13 & 43.88$\pm$0.21 & 2.48$\pm$0.51 & 44.49$\pm$0.13 \\
TBabs*(zpo+compTT) & 	0.94$\pm$0.09 & 44.09$\pm$0.10 & 0.62$\pm$0.10 & 43.88$\pm$0.19 & 2.49$\pm$0.25 & 44.50$\pm$0.10 \\  
TBabs*cutoffpl & 	0.93$\pm$0.17 & 44.08$\pm$0.19 & 0.59$\pm$0.11 & 43.86$\pm$0.19 & 2.34$\pm$0.44 & 44.47$\pm$0.19 \\
TBabs*(bknpo+zgauss) & 	0.95$\pm$0.19 & 44.09$\pm$0.20 & 0.63$\pm$0.13 & 43.88$\pm$0.17 & 2.62$\pm$0.52 & 44.51$\pm$0.19 \\
TBabs*(zpo+zbbody) & 	0.94$\pm$0.02 & 44.09$\pm$0.02 & 0.62$\pm$0.01 & 43.88$\pm$0.02 & 2.49$\pm$0.05 & 44.50$\pm$0.02 \\
TBabs*bknpo*zxipcf & 	0.95$\pm$0.13 & 44.06$\pm$0.14 & 0.63$\pm$0.09 & 43.86$\pm$0.14 & 2.59$\pm$0.36 & 44.48$\pm$0.14 \\ 
\hline\hline
\end{tabular}
\tablefoot{Columns: (1) model; (2) observed flux between 0.3-10 keV in units of $10^{-12}$ erg s$^{-1}$ cm$^{-2}$; (3) logarithm of the intrinsic luminosity (erg s$^{-1}$), corrected for Galactic absorption, between 0.3-10 keV; (4) observed flux 2-10 keV; (5) logarithm of the intrinsic luminosity 2-10 keV; (6) observed flux 0.3-70 keV; (7) logarithm of the intrinsic luminosity 0.3-70 keV.}
\end{table*}

\end{appendix}

\end{document}